\documentclass[11pt]{article}
\usepackage{amsmath,amssymb,color,graphics,epsfig,cite}

\usepackage{amsfonts}

\newcommand{\be}{\begin{equation}}
\newcommand{\ee}{\end{equation}}
\newcommand{\bea}{\setlength\arraycolsep{2pt} \begin{eqnarray}}
\newcommand{\eea}{\end{eqnarray}}
\newcommand{\nn}{\nonumber}

\def\ft#1#2{{\textstyle{\frac{\scriptstyle #1}{\scriptstyle #2} } }}
\def\fft#1#2{{\frac{#1}{#2}}}

\def\0{{\sst{(0)}}}
\def\1{{\sst{(1)}}}
\def\2{{\sst{(2)}}}
\def\3{{\sst{(3)}}}
\def\4{{\sst{(4)}}}
\def\5{{\sst{(5)}}}
\def\6{{\sst{(6)}}}
\def\7{{\sst{(7)}}}
\def\8{{\sst{(8)}}}
\def\sst#1{{\scriptscriptstyle #1}}

\begin{document}

\begin{center}
{\Large {\bf A New Approach to Black Hole Thermodynamics\\ in General EMD-like Theories}}

\vspace{20pt}

Peng-Yu Wu and H. L\"{u}

\vspace{10pt}

{\it Center for Joint Quantum Studies, Department of Physics,\\
School of Science, Tianjin University, Tianjin 300350, China }

\vspace{40pt}

\underline{ABSTRACT}
\end{center}

We consider general EMD-like theories involving arbitrary numbers of both Maxwell fields and dilatonic scalars. We study charged black holes and derive several algebraic and differential relations among the thermodynamic quantities, scalar charges and the non-extremal parameter. These allow us to obtain two sets of master equations of thermodynamic quantities, depending on the choice of basic variables parameterizing the system. Solutions to either set of the master equations give rise to all the thermodynamic quantities, without constructing black hole solutions. We obtain all the thermodynamic quantities perturbatively in the power series expansions of charges, which we then verify using perturbative black hole solutions up to and including the quartic order in charges.

\vfill{wupy2023@tju.edu.cn\ \ \  mrhonglu@gmail.com}


\thispagestyle{empty}
\pagebreak

\tableofcontents
\addtocontents{toc}{\protect\setcounter{tocdepth}{2}}

\newpage

\section{Introduction}

Although we are still lacking a generally accepted consistent quantum theory of gravity, the study of quantum effects using a semiclassical approach has important applications in General Relativity, including early cosmology and black hole physics. In the latter research area, Hawking's derivation \cite{Hawking:1974rv} of black hole radiation based on the semiclassical method has since promoted the black hole dynamics  \cite{Bardeen:1973gs} to thermodynamics. By now, the thermodynamic relations are among the most robust probes of black hole solution spaces. In conventional applications, one first constructs black hole solutions by solving the gravitational field equations, and then analyses the horizon and asymptotic geometric structures, and finally evaluates the mass, entropy, temperature and other relevant thermodynamic quantities.  This order is natural when an exact metric is available, but it becomes cumbersome and restrictive in theories where exact solutions may not exist.

We can read off the thermodynamic quantities locally from either the asymptotic or the near-horizon geometries. For example, for the asymptotically-Minkowski spacetime, the mass can be locally determined from the asymptotic geometry. The temperature can be determined from the near-horizon geometry by requiring the proper period of Euclidean time to avoid a conic singularity \cite{Gibbons:1976ue}. However, we need the full black hole solution to relate the asymptotic data to the horizon data. On the other hand, the thermodynamic properties of a black hole are completely determined by the corresponding gravitational theory. It is thus natural to ask whether one can derive the thermodynamics of a black hole in a given theory, bypassing the step of constructing the solution altogether. Indeed the first law of black hole thermodynamics, which would appear miraculous, was established as a direct mathematical consequence of the Iyer-Wald identity \cite{Wald:1993nt,Iyer:1994ys} based on the covariant phase-space formalism. This is a concrete example of establishing one important black hole thermodynamic property without needing to construct a specific black hole solution. Recently, Reall-Santos \cite{Reall:2019sah} established that leading-order curvature correction to black hole thermodynamics in Einstein gravity can be obtained without constructing the perturbative solution, provided that the zeroth order solution is known. (See also \cite{Hu:2023gru,Xiao:2023two} for dealing with subtleties in asymptotic-anti-de Sitter black holes.) We therefore have seen a continuous effort to derive the black hole thermodynamics directly from the theories; however, black hole solutions are still required to investigate the full thermodynamic space.

Recently, it was proposed \cite{Lu:2025eub,Yang:2025rud} that for some special class of string-inspired theories, black hole thermodynamics can be completely determined without constructing a local solution. The theories considered in \cite{Lu:2025eub,Yang:2025rud} are Einstein-Maxwell-dilaton (EMD) theory, as well as the Einstein-Maxwell-Maxwell-dilaton (EMMD) generalizations. For suitable dilaton coupling constants, exact solutions were known in the literature, associated with Toda equations of various Lie groups \cite{Lu:1996hh, Lu:2013uia,Ivashchuk:2002ge,Ivashchuk:2013jja,Lu:2026lkv}. These exact solutions provide important checks of the proposal. Furthermore, this technique was also generalized to derive black $p$-brane thermodynamics in diverse dimensions \cite{Han:2026bim}. This paper is an expanded version of the letter \cite{Yang:2025rud}, focusing on two major extensions:
\begin{itemize}

\item Consider the complete general class of EMD-like theories, involving arbitrary numbers of both Maxwell fields and dilatonic scalars.

\item Provide concrete proofs to all the criteria used in the proposal of \cite{Yang:2025rud}, and also provide a somewhat new technique to determine the thermodynamics.

\end{itemize}
We derive, based on the properties of the theories, several algebraic and differential relations among the thermodynamic quantities. These allow us to determine two sets of master equations of thermodynamic quantities, depending on the choice of the basic variables of parameterizing the thermodynamic quantities. One master equation is simply the extension of \cite{Yang:2025rud} for the more general EMD-like theories. The other master equation is new, using directly thermodynamic variables in microcanonical ensemble. The solutions to either of the master equations can yield all the thermodynamic quantities without constructing black hole solutions.

The paper is organized as follows. In Section \ref{sec:theory}, we introduce the general EMD-like theories and the corresponding equations of motion for spherically-symmetric and static charged black hole ansatz. We restrict ourselves to solutions that are asymptotic to Minkowski spacetimes and have an event horizon. In Section \ref{sec:algebraicrelations}, we analyse the asymptotic falloffs and horizon geometry and obtain several algebraic relations, including the long-range force identity,  scalar charge identity and the Smarr relation. In Section \ref{sec:diffrelations}, we derive two differential relations: the first law and a scalar-charge differential relation. In Section \ref{sec:twomaster}, depending on the choice of basic variables used for thermodynamic quantities, we derive two sets of master equations that determine all the thermodynamic quantities. In other words, solutions to either set give rise to all the thermodynamic quantities. In Section \ref{sec:perturbation}, we solve the master equations perturbative in terms of power series expansion of the electric charges, and show that the resulting thermodynamic quantities are indeed precisely the same as those obtained from constructing explicit perturbative black hole solutions. In Section \ref{sec:numerical}, we (randomly) choose a specific theory and apply a numerical method to demonstrate the correctness of the master equations to derive the thermodynamic quantities.  We conclude our paper in Section \ref{sec:conclusion}

\section{EMD-like theories and black hole equations of motion}
\label{sec:theory}

\subsection{Theories, ansatz and equations}
In this section, we consider a general class of EMD-like theories involving multiple numbers of both Maxwell fields and dilatonic scalars. The bulk action is
\begin{equation}
I=\fft{1}{16\pi} \int d^D x\,\sqrt{-g}\left[
R-\frac12\partial_\mu\vec\phi\cdot\partial^\mu\vec\phi
-\frac14\sum_{i=1}^{N_A}e^{\vec a_i\cdot\vec\phi}\, F_i^2
\right]\,,\qquad F_i = dA_i\,,
\label{eq:emmd-action}
\end{equation}
where $\vec \phi = (\phi_1, \phi_2\, \cdots \phi_{N_\phi})$ is the dilaton vector and $\vec a_i$ are the constant vectors describing the non-minimal coupling between the dilatons and the Maxwell field strength $F_i$. Such an action is typically considered in $p$-brane constructions in string theories, with the Maxwell fields replaced by generic form fields, e.g.~\cite{Duff:1996hp}. We assume that the number of the dilatons $N_\phi$ is smaller than the total number of Maxwell fields $N_A$, since if the former is bigger than the latter, we can choose a basis so that the superfluous dilatons become minimally coupled massless scalars that are irrelevant in black hole construction and they should be zero onset.

The covariant equations of motion associated with the variations of the dilatons, the Maxwell potentials and the metric are respectively given by
\bea
&&\Box \vec \phi = \fft14 \sum_{i=1}^{N_A}\vec a_i\, e^{\vec a_i \cdot \vec \phi} F_i^2\,,\qquad d \big(e^{\vec a_i \cdot \vec \phi} {*F_i}\big) = 0\,,\nn\\
&&G_{\mu\nu}=\fft12\Big(\partial_{\mu}\vec \phi \cdot  \partial_{\nu}\vec \phi -
\fft12 \partial_\rho\vec\phi\cdot\partial^\rho\vec\phi\, g_{\mu\nu}\Big) +
\fft12 \sum_{i=1}^{N_A} e^{\vec a_i \cdot \vec \phi} \Big((F_i^2)_{\mu\nu} - \fft14
F_i^2\, g_{\mu\nu}\Big)\,.
\eea
In this paper, we consider electrically-charged spherically-symmetric and static black holes. We adopt the black $p$-brane type of ansatz \cite{Duff:1996hp}, given by
\begin{equation}
ds^2=-e^{-2U(r)}h(r)\,dt^2 +e^{2U(r)/(D-3)}\left(\frac{dr^2}{h(r)}+r^2d\Omega_{D-2}^2
\right)\,,\qquad \vec\phi=\vec\phi(r)\,, \label{eq:U-ansatz}
\end{equation}
where $h(r)$ is the blackening factor, given by
\be
h(r) = 1 - \fft{2\mu}{r^{D-3}}\,.
\ee
The main advantage of this $p$-brane like ansatz is that the horizon is solely determined by the blackening factor $h(r)$, with $\mu$ referred to as the blackening or non-extremal parameter. Such simplification is possible for the EMD-like theories that are inspired by string theories. The Maxwell equations can be easily solved, given by
\begin{equation}
F_i= \frac{4Q_i}{r^{D-2}} \exp\left(-2U-\vec a_i\cdot\vec\phi\right)\, dt\wedge dr\,,
\label{eq:maxwell-first-integral}
\end{equation}
where the integration constants $Q_i$ parameterize the electric charges, defined by
\be
Q_i^e = \fft{1}{16\pi} \int e^{\vec a_i\cdot \phi} {*F}_i = \fft{\Omega_{D-2}}{4\pi} Q_i\,.
\label{eq:electriccharge}
\ee
Here $\Omega_{D-2}$ is the volume of the unit $(D-2)$-sphere. For later purposes, we also define parameters $(\delta, \kappa)$, all given by
\be
\delta \equiv \fft{2(D-3)}{D-2}\,,\qquad
\kappa\equiv \fft{(D-2)\Omega_{D-2}}{8\pi}\,,\qquad \Omega_{D-2} \equiv \frac{2\, \pi^{\frac{D-1}{2}}}{\Gamma(\frac{D-1}{2})}\,.
\label{eq:deltakappa}
\ee
To describe the equations of motion for the metric function $U(r)$ and the dilatons $\vec \phi(r)$, it is useful to introduce the linear second-order radial differential operator
\begin{equation}
\mathcal D[X] \equiv X'' +\frac{D-3+h}{rh}X' = X'' +\left(\frac{D-2}{r}
+\frac{h'}{h} \right)X'.
\label{eq:radial-operator}
\end{equation}
After substituting the ansatz into the Einstein and scalar equations, we find that the functions $U(r)$ and $\vec\phi(r)$ satisfy
\begin{equation}
\mathcal D[U]+\delta\sum_{i=1}^{N_A}S_i=0\,,\qquad
\mathcal D[\vec \phi] +2\sum_{i=1}^{N_A}\vec a_i S_i=0\,,\label{eq:Uphi-eom}
\end{equation}
together with the first-order Hamiltonian constraint
\begin{equation}
\mathcal H\equiv \frac{U'^2}{\delta} +\frac14\vec\phi'^{\,2}-\frac{(D-2)\mu}{r^{D-2}f}U'
-\sum_{i=1}^{N_A}S_i=0.
\label{eq:hamiltonian-U}
\end{equation}
Here the ``source'' terms $S_i$'s are given by
\begin{equation}
S_i(r)\equiv\frac{\chi_i(r)}{h(r)}\,,\qquad \chi_i(r)\equiv
\frac{4Q_i^2}{r^{2(D-2)}}
\exp\left(-2U-\vec a_i\cdot\vec\phi\right)\,.\label{eq:chi-S}
\end{equation}
Note that the first-order Hamiltonian constraint is consistent since applying \eqref{eq:Uphi-eom}, we obtain
\begin{equation}
\mathcal H'=-2\left(\frac{D-2}{r}+\frac{h'}{h}\right)\mathcal H\,,
\label{eq:hamiltonian-propagation}
\end{equation}
which in itself implies more generally that
\be
{\cal H}(r) = \fft{C}{r^{2(D-2)} h(r)^2}\,.\label{eq:hamiltonian-first-integral}
\ee
Einstein theory is special in that it forces the vanishing integration constant, i.e.~ $C=0$.

Closed-form solutions of Eqs.~\eqref{eq:Uphi-eom} and \eqref{eq:hamiltonian-U} are
known only for special choices of the dilaton-coupling vectors. Indeed, many such solutions were constructed in literature. It was observed that the equations can be cast into one-dimensional Toda like equations and become Toda equations of any Lie group for suitable choices of $\vec a_i$ \cite{Lu:1996hh,Lu:2013uia,Ivashchuk:2002ge,Ivashchuk:2013jja, Lu:2026lkv}. This significantly enlarges the spectrum of exact solutions. However, exact solutions are generally lacking for generic $\vec a_i$.

\subsection{Horizon smoothness and propagation of the Hamiltonian constraint}\label{hamiltonian-regularity}

The most general solution that is asymptotic to Minkowski spacetime may not describe a black hole, but a solution involving a naked singularity. For a black hole solution, the outer horizon is necessarily located at the root of $h(r)$. Let $r=r_h$ be a non-extremal horizon, we have
\begin{equation}
\mu = \ft12 r_h^{D-3}\,,\qquad f(r_h)=0\,,\qquad f'(r_h)\neq0\,. \label{eq:horizon-definition}
\end{equation}
For the black hole branch satisfying the smooth nonsingular event-horizon conditions, we impose
\begin{equation}
U(r)=U_h+U_1(r-r_h)+O\left((r-r_h)^2\right)\,,\qquad
\vec\phi(r)=\vec\phi_h+\vec\phi_1(r-r_h)+O\left((r-r_h)^2\right).
\label{eq:regular-Uphi}
\end{equation}
Multiplying the $U$ equation in \eqref{eq:Uphi-eom} by $h(r)$ and taking $r\to r_h$ gives the horizon smoothness condition
\begin{equation}
h'_hU'_h+\delta\sum_{i=1}^{N_A}\chi_{i,h}=0\,.
\label{eq:U-horizon-regularity}
\end{equation}
On the other hand, the Hamiltonian can be rearranged as
\begin{equation}
\mathcal H=\frac{U'^2}{\delta}+\frac14\vec\phi'^{\,2}-\frac1{h}
\left(\frac{h'}{\delta}U'+\sum_{i=1}^{N_A}\chi_i\right).
\label{eq:hamiltonian-pole-split}
\end{equation}
Equation \eqref{eq:U-horizon-regularity} sets the coefficient of the apparent $1/h$ pole to zero on the horizon. Thus $\mathcal H$ is finite for the branch satisfying the horizon smoothness conditions. However, \eqref{eq:hamiltonian-first-integral} would behave as $1/h^2$ near the horizon if $C\neq0$. Smoothness on the horizon therefore requires $C=0$
and hence we have $\mathcal H(r)=0$ everywhere. However, this implication is only one-way. The Hamiltonian constraint alone does not guarantee horizon smoothness. In particular, there exist solutions where $U$ diverges at $r=r_h$, so that $r=r_h$ ceases to be an horizon, but a naked singularity.  Thus, requiring the metric and scalar fields to remain finite at a non-extremal horizon selects the branch satisfying the smooth non-degenerate event-horizon conditions.

\section{Algebraic thermodynamic identities}
\label{sec:algebraicrelations}

In this paper, we focus on the electrically-charged spherically-symmetric and static black holes in the EMD-like theories \eqref{eq:emmd-action}. The general solutions can be described by a variety of quantities, including mass $M$, electric charges $Q_i^e$, scalar charges $\vec \Sigma$, temperature $T$, entropy $S$ and electric potential $\Phi_i$. These quantities are not all independent. Owing to special properties of the EMD-like theories, some relations are purely algebraic and we present them here, without constructing exact or numerical solutions.

\subsection{The quadratic force relation from the Hamiltonian constraint}

The long-range force between two identical black holes depends only on the asymptotic properties of the spacetime. It was initially studied in \cite{Cremonini:2023vwf, Cremonini:2024eog} for extremal black holes in EMD or EMMD theories and was later generalized to non-extremal ones \cite{Lu:2025eub,Yang:2025rud}. In fact, the spacetime does not even have to be a black hole \cite{Lu:2025eub,Yang:2025rud}. Here, we present a derivation of the quadratic force relation from the Hamiltonian constraint.

Since the long-range force involves only the asymptotic parameters, we begin by analysing the asymptotic Minkowski falloffs of the equations \eqref{eq:Uphi-eom} and \eqref{eq:hamiltonian-U}. For generality, we keep the asymptotic dilaton moduli arbitrary, thus the Minkowski spacetime is described by
\be
U(\infty) =0\,,\qquad \vec \phi(\infty)=\vec \lambda\,.\label{asymptoticboundary}
\ee
Since the $U$ and $\vec\phi$ equations can be written in the divergence forms as
\begin{equation}
\left(r^{D-2} h U'\right)'
= -\frac{4\delta}{r^{D-2}}\sum_{i=1}^{N_A} Q_i^2e^{-2U-\vec a_i\cdot\vec\phi}\,,\qquad
\left(r^{D-2} h {\vec \phi}^{\prime}\right)'=-\frac{8}{r^{D-2}}\sum_{i=1}^{N_A}
\vec a_i Q_i^2 e^{-2U-\vec a_i\cdot\vec\phi}\,,
\label{eq:Uphi-divergence-form}
\end{equation}
it follows from \eqref{asymptoticboundary} that after the first integration, we have
\begin{equation}
r^{D-2} h U' = C_U+O\left(r^{-(D-3)}\right)\,, \qquad r^{D-2} h \vec\phi' =
\vec C_\phi+O\left(r^{-(D-3)}\right)\,.
\label{eq:asymptotic-first-integrals}
\end{equation}
Using $h=1+O\left(r^{-(D-3)}\right)$, we obtain
\begin{equation}
U'= \frac{C_U}{r^{D-2}} +O\left(r^{-(2D-5)}\right),
\qquad
\vec\phi'=\frac{\vec C_\phi}{r^{D-2}}
+O\left(r^{-(2D-5)}\right).
\label{eq:asymptotic-derivative-falloffs}
\end{equation}
A further integration gives the universal massless falloffs of order $1/r^{D-3}$:
\begin{equation}
U(r)=\frac{u}{r^{D-3}}+O\left(r^{-2(D-3)}\right)\,,\qquad
\vec\phi(r)= \vec\lambda+ \frac{2(D-2)}{(D-3)\kappa} \frac{\vec\Sigma}{r^{D-3}}
+O\left(r^{-2(D-3)}\right).
\label{eq:Uphi-asymptotic}
\end{equation}
where
\begin{equation}
u=-\frac{C_U}{D-3},
\qquad
\vec\Sigma=-\frac{\Omega_{D-2}}{16\pi}\vec C_\phi=
-\frac{\kappa}{2(D-2)}\vec C_\phi.
\label{eq:u-and-scalar-charge}
\end{equation}
Thus, the $r^{-(D-3)}$ behavior is the nonconstant radial harmonic falloff of massless fields in an asymptotically-flat $D$-dimensional spacetime.  If the corresponding coefficient vanishes, the first nonzero correction may start at a higher order. However, in our two-dervative EMD-like theories, nontrivial asymptotic parameters must appear in the leading $1/r^{D-3}$ falloffs; otherwise, the solutions all reduce to the Minkowski vacuum.

The vector parameter $\vec \Sigma$ can be viewed as the asymptotic scalar charge, defined by
\be
\vec \Sigma = \fft{1}{16\pi} \int_{r\rightarrow \infty } \iota_{\partial_t}(*d\phi)\,,
\label{scalarcharge}
\ee
where \(\iota_{\partial_t}\) denotes contraction with Killing vector $\partial_t$. Substituting the asymptotic expansion \eqref{eq:Uphi-asymptotic} into the Hamiltonian constraint \eqref{eq:hamiltonian-U}, we have
\begin{equation}
\kappa^2\left(u^2+2\mu u\right) + \frac{4}{\delta} \left[
\vec\Sigma\cdot\vec\Sigma -\sum_{i=1}^{N_A} e^{-\vec a_i\cdot\vec\lambda}(Q_i^e)^2\right]
=0\,.
\label{eq:quadratic-u-relation}
\end{equation}
The parameter $u$ is related to the ADM mass. To see this, we note that
\be
g_{tt} = - e^{-2U} f \sim -1 + \fft{2(\mu+u)}{r^{D-3}} + \cdots\,.
\ee
Therefore, we have
\be
M=\kappa (\mu + u)\,.\label{eq:mass-u-relation}
\ee
Thus, the identity \eqref{eq:quadratic-u-relation} becomes the quadratic constraint on the mass, electric charges, scalar charge, as well as the non-extremal parameter:
\begin{equation}
M^2+\frac{4}{\delta}\left[\vec\Sigma\cdot\vec\Sigma
-\sum_{i=1}^{N_A}e^{-\vec a_i\cdot\vec\lambda}(Q_i^e)^2\right]
=\kappa^2\mu^2\,.
\label{eq:quadratic-force-relation}
\end{equation}
Following the definition of the long-range force mediated by massless fields, we find that
the combination on the left-hand side is precisely the coefficient of
the long-range force between two identical asymptotically-flat
configurations:
\begin{equation}
-\lim_{r\rightarrow \infty} r^{D-2}\,{\rm  Force} = M^2+ \frac{4}{\delta} \left[
\vec\Sigma\cdot\vec\Sigma -\sum_{i=1}^{N_A}
e^{-\vec a_i\cdot\vec\lambda}(Q_i^e)^2 \right]\,.
\label{eq:long-range-force}
\end{equation}
In this definition of the inverse-squared long-range force law, the gravitation mass and scalar charges contribute attractive force while the Maxwell charges contribute the repulsive force. Consequently, the Hamiltonian constraint gives
\begin{equation}
\lim_{r\rightarrow \infty} r^{D-2}\,{\rm  Force} = -\kappa^2\mu^2\,. \label{eq:force-mu-relation}
\end{equation}
This relation is independent of whether the spacetime is a black hole or it has a naked singularity. If the solution does describe a black hole, where the horizon $r_h$ is located at $h(r_h)=0$, the blackening parameter $\mu$ therefore has an invariant interpretation: it measures the residual attractive force after the electric and scalar field exchanges are included.  In other words, for a non-extremal solution with $\mu>0$, the force is therefore
attractive in the convention of \eqref{eq:long-range-force}.  In the extremal limit $\mu=0$, the gravitational and scalar attractions are exactly balanced at the long range by the electric repulsion, and the long-range force vanishes. This extremal relation was obtained in \cite{Cremonini:2023vwf,Cremonini:2024eog, Cremonini:2026lmz} for EMD and EMMD theories. This balancing is weaker than the no-force condition in some supersymmetric black holes such as the extremal Reissner-Nordstr\"om black hole, where electric repulsion and gravitational attraction balance out precisely in any separations. The relation \eqref{eq:force-mu-relation} supplies the first boundary input for the direct thermodynamic construction and has been obtained without the full radial profiles.

Notice that the expression $e^{-\vec a_i\cdot\vec\lambda}(Q_i^e)^2$ in the long-range force formula is invariant under a constant dilaton shift, namely
\begin{equation}
\vec\lambda\longrightarrow\vec\lambda+\vec c\,,
\qquad Q_i^e\longrightarrow
e^{\frac12\vec a_i\cdot\vec c}Q_i^e\,.
\label{eq:dilaton-shift-charge-transformation}
\end{equation}
Thus \eqref{eq:quadratic-force-relation} is independent of the representative chosen on the constant-dilaton-shift orbit.  If we introduce the shift-invariant dressed charges
\begin{equation}
\widehat Q_i^e\equiv e^{-\frac12\vec a_i\cdot\vec\lambda}Q_i^e,
\label{eq:dressed-electric-charges}
\end{equation}
then \eqref{eq:quadratic-force-relation} takes the simpler form
\begin{equation}
M^2+ \frac{4}{\delta}\left( \vec\Sigma\cdot\vec\Sigma
-\sum_{i=1}^{N_A}(\widehat Q_i^e)^2
\right) = \kappa^2\mu^2\,.
\label{eq:quadratic-force-relation-dressed}
\end{equation}
It is this form that was proposed in \cite{Lu:2025eub,Yang:2025rud} in four dimensions and in \cite{Lu:2026lkv} for general dimensions.

\subsection{An algebraic scalar-charge identity}

In the previous subsection, we have seen that the general solutions that are asymptotic to Minkowski spacetime are parameterized by $(M, Q_i^e, \vec \Sigma)$. If such a solution actually describes a black hole, the scalar $\vec \Sigma$ are not independent parameters, but function of $(M, Q_i^e)$. This weaker version of the no-scalar-hair conjecture, proposed in \cite{Lu:2025eub}, can be concretely established in the general EMD-like theories. To do so, we first establish an algebraic relation for $\vec \Sigma$. We derive this relation using two methods.

\subsubsection{Integrated radial equations and horizon potentials}\label{sec:integrated-identities}

The long-range force relation is a purely asymptotic property, and therefore it does not specify how the asymptotic charges enter the thermodynamic quantities associated with the horizon.  To obtain the missing information, we impose the smooth non-degenerate horizon conditions on the outer horizon and integrate the $U$ and scalar equations over the exterior region.

The quadratic relation \eqref{eq:quadratic-force-relation} follows
only from the Hamiltonian constraint and asymptotic flatness; it does
not require horizon smoothness.  We now impose the additional
assumption that the solution has a smooth event horizon at
$r=r_h$, with $\mu=\fft12 r_h^{D-3}$ and all the quantities $(U(r_h),\ \vec\phi(r_h),\ U'(r_h),\ \vec\phi'(r_h))$ are finite. Under this assumption, we have
\begin{equation}
\left.r^{D-2} h U'\right|_{r=r_h}=0\,,
\qquad
\left.r^{D-2} h \vec\phi'\right|_{r=r_h}=0\,.
\label{eq:regular-horizon-boundary-terms}
\end{equation}
For each Maxwell field \eqref{eq:maxwell-first-integral}, the thermodynamic electric potential is the gauge-invariant potential difference.  It connects the local Maxwell field to the intensive variable entering the first law, namely
\begin{equation}
\Phi_i \equiv A_{it}(r_h)-A_{it}(\infty) = 4 Q_i I_i\,,\qquad
I_i \equiv \int_{r_h}^{\infty} \frac{dr}{r^{D-2}}
e^{-2U-\vec a_i\cdot\vec\phi}\,.
\label{eq:potential-difference}
\end{equation}
On the other hand, integrating both equations in \eqref{eq:Uphi-divergence-form} from the horizon to infinity gives
\begin{equation}
C_U = -4\delta\sum_{i=1}^{N_A}Q_i^2 I_i\,,
\qquad
\vec C_\phi = -8\sum_{i=1}^{N_A}\vec a_iQ_i^2 I_i\,.
\label{eq:CU-Cphi-integral-relations}
\end{equation}
Substituting \eqref{eq:potential-difference} into the above integrated $U$ and scalar equations then yields the following identities:
\begin{equation}
C_U=-\delta\sum_{i=1}^{N_A}Q_i\Phi_i\,, \qquad
\vec C_\phi = -2\sum_{i=1}^{N_A}\vec a_iQ_i\Phi_i\,.
\label{eq:CU-Cphi-potential-relations}
\end{equation}
Using \eqref{eq:electriccharge}, \eqref{eq:u-and-scalar-charge},
and \eqref{eq:mass-u-relation}, we obtain two algebraic horizon-asymptotic relations
\begin{equation}
M=\kappa\mu+\sum_{i=1}^{N_A}Q_i^e\Phi_i\,,
\label{eq:mass-potential-identity}
\end{equation}
and
\begin{equation}
\vec\Sigma = \frac12\sum_{i=1}^{N_A} \vec a_iQ_i^e\Phi_i\,.
\label{eq:scalar-charge-potential-identity}
\end{equation}
The first relation separates the non-extremal contribution $\kappa\mu$ from the electrostatic work terms.  The second is the scalar-charge identity central to this work: on the branch satisfying the horizon smoothness conditions, the scalar-charge vector is fixed by the electric potentials weighted by the coupling vectors and is not an independent thermodynamic variable.  Both statements follow directly from the radial equations. The scalar-charge relation for the electrically-charged black hole in  EMD theory was previously obtained in \cite{Pacilio:2018gom}.

\subsubsection{Noether interpretation of the scalar-charge identity}\label{sec:noether-interpretation}

The scalar-charge identity \eqref{eq:scalar-charge-potential-identity} also admits a covariant interpretation.  The action is invariant under a constant dilaton shift accompanied by a compensating rescaling of the Maxwell potentials, namely
\begin{equation}
\vec\phi\longrightarrow\vec\phi+\vec c\,,
\qquad
A_i\longrightarrow e^{-\frac12\vec a_i\cdot\vec c}A_i\,.
\label{eq:dilaton-shift-main}
\end{equation}
The associated on-shell Noether current is
\begin{equation}
\vec J^{\,\mu}=-\nabla^\mu\vec\phi+\frac12\sum_i\vec a_i e^{\vec a_i\cdot\vec\phi}F_i^{\mu\nu}A_{i\nu}\,,
\qquad \nabla_\mu\vec J^{\,\mu}=0\,.
\label{eq:noether-current-main}
\end{equation}
For static, spherically symmetric ansatz, this implies that $\partial_r(\sqrt{-g} \vec J^r)=0$, and hence the integral
\begin{equation}
\vec{\mathcal I}(r) = \int_{S^{D-2}} d\Omega_{D-2}\, \sqrt{-g}\,\vec J^{\,r}
\end{equation}
is independent of $r$. For our purely electric ansatz, we have
\begin{equation}
\vec J^{\,r} =-g^{rr}\vec\phi\,' +\frac{1}{2} \sum_i\vec a_i e^{\vec a_i\cdot\vec\phi}
  F_i^{rt}A_{it}\,.
\end{equation}
Therefore, the $r$-independent integral becomes
\bea
\vec{\mathcal I}(r) &=&  -\int_{S^{D-2}}d\Omega_{D-2}\,
\sqrt{-g}\,g^{rr}\vec\phi\,' + \frac{1}{2} \sum_i \vec a_i A_{it}(r) \int_{S^{D-2}}d\Omega_{D-2}\, \sqrt{-g}\, e^{\vec a_i\cdot\vec\phi}F_i^{rt}\nn\\
&=&-\int_{S^{D-2}}d\Omega_{D-2}\, \sqrt{-g}\,g^{rr}\vec\phi\,' + 8\pi
\sum_i \vec a_i Q_i^e A_{it}(r)\,.
\eea
Choosing the gauge $ A_{it}(\infty)=0$, the asymptotic infinity \eqref{eq:Uphi-asymptotic} implies that $\vec{\mathcal I}(\infty)=16\pi\vec\Sigma$. On the smooth horizon, $g^{rr}=0$ and the scalar field remains finite, we have $\vec{\mathcal I}(r_h)=8\pi \sum_i \vec a_i Q_i^e \Phi_i$. Equating $\vec{\mathcal I}(\infty)=\vec{\mathcal I}(r_h)$ leads to \eqref{eq:scalar-charge-potential-identity}. The agreement with \eqref{eq:scalar-charge-potential-identity} shows that the scalar-charge relation is not tied to a particular radial parametrization; it is the boundary realization of the global dilaton-shift symmetry. Note, however, the identity \eqref{eq:scalar-charge-potential-identity} in itself does not imply that the scalar $\vec \Sigma$ must depend on $(M,Q_i^e)$, as the identity involves new quantities such as the electric potentials $\Phi_i$. The proof that $\vec \Sigma= \vec \Sigma (M, Q_i^e)$ for black holes will come in Section \ref{sec:diffrelations}.

\subsection{Asymptotic-horizon relation and Smarr formula}\label{sec:smarr}

For the metric ansatz \eqref{eq:U-ansatz} with regular horizon located at $h(r_h)=0$, the temperature and entropy can be calculated in the standard method, given by
\be
T=\fft{D-3}{4\pi r_h} \exp\big(-\frac{D-2}{D-3} U(r_h)\big)\,,\qquad
S=\frac14 r_h^{D-2} \exp\big(\frac{D-2}{D-3} U(r_h)\big)\, \Omega_{D-2}\,.\label{eq:TSdef}
\ee
It follows that the product of temperature and entropy depends only on $\mu$:
\be
T S = \frac{D-3}{D-2} \kappa\,\mu=\frac{\delta\kappa}{2}\mu\,.\label{eq:TSmu}
\ee
This horizon-asymptotic relation was first observed in \cite{Lu:2025eub,Yang:2025rud} for four-dimensional EMD and EMMD black holes. Combining \eqref{eq:TSmu} with \eqref{eq:mass-potential-identity}, we obtain
\begin{equation}
M =\frac{D-2}{D-3}TS +
\sum_{i=1}^{N_A}\Phi_iQ_i^e\,.\label{eq:Smarr-relation}
\end{equation}
This is the usual Smarr relation, which can also be derived from the generalized Komar integration of the timelike Killing vector $\xi=\partial/\partial_t$.

\section{Differential thermodynamic identities}
\label{sec:diffrelations}

\subsection{The first law}

The first law of black hole thermodynamics can be established by the Iyer-Wald formalism, without explicit black hole solutions. It turns out that the scalar moduli $\vec \lambda =\vec \phi(\infty)$ can also contribute to the first law, and its thermodynamic conjugate is the scalar charge vector $\vec \Sigma$ \cite{Gibbons:1996af}. The extended first law for the electrically-charged black holes is given by
\begin{equation}
dM= T\,dS + \sum_{i=1}^{N_A}\Phi_i\,dQ_i^e - \vec\Sigma\cdot d\vec\lambda\,.
\label{eq:extended-first-law}
\end{equation}
As was discussed, the scalar modulus vector $\vec \lambda$ can be simply set to zero by the constant shift symmetry of the dilatonic scalars such that the scalar charge vector $\Sigma$ plays no essential role in thermodynamics. However, keeping $\vec\lambda$ generic makes the shifting symmetry manifest in the first law, which is essential for deriving an important differential relation of the scalar charges later. The first law \eqref{eq:extended-first-law} is the consequence of the radial conservation of the variation of the Hamiltonian associated with the timelike Killing vector $\xi=\partial_t$ in the Iyer-Wald formalism, namely $\delta {\cal H}_\xi(r_h)=\delta {\cal H}_\xi(\infty)$, where
\bea
\delta {\cal H}_\xi(r) &=& -\fft{(D-2) \Omega_{D-2}}{16\pi (D-3)} r^{D-2}\Big(
(U' + \fft{D-3}{r}) \delta f + 2(U'\delta U + \delta U') f - f' \delta U\Big)\nn\\
&&+ \sum_{i} A_{t,i}(r) \delta Q_i^e - \fft{\Omega_{D-2}}{16\pi} r^{D-2}\, f\, \vec\phi' \cdot \delta\vec \phi\,.\label{eq:wald}
\eea
Note that here $\delta$ denotes the variation of the on-shell integration constants of the black hole solution; there should be no confusion with the parameter $\delta$ introduced in this paper. In particular, the last term in \eqref{eq:wald} vanishes identically on the horizon, but it contributes the last term in the extended first law \eqref{eq:extended-first-law}.

\subsection{Scalar-charge differential identity}

The weak version of the no-scalar-hair conjecture states that if we take the scalar moduli $\vec\lambda=0$, the scalar charge vector $\vec \Sigma$ is not independent, but a vector function of mass $M$ and electric charges $Q_i^e$'s. The quadratic force relation \eqref{eq:quadratic-force-relation} implies that we can use $(\mu, Q_i^e)$ as basic variables and then scalar charges must be certain specific functions of these variables, namely $\vec\Sigma=\vec\Sigma(\mu, Q_i^e)$. Although we already derived an algebraic relation \eqref{eq:scalar-charge-potential-identity} for the scalar charge, it involves additional electric potentials $\Phi_i$ so that it is not apparent whether $\vec \Sigma$ is independent. In order to determine the vector function $\vec\Sigma=\vec\Sigma(\mu, Q_i^e)$, it was proposed in \cite{Lu:2025eub,Yang:2025rud,Lu:2026lkv} that the scalar charges satisfy the differential relation
\be
\vec \Sigma =\mu \fft{\partial\vec \Sigma}{\partial\mu}+\frac12 \sum_{i=1}^{N_A} \vec a_i Q_i^e \fft{\partial M}{\partial Q_i^e}\,.\label{eq:scalarchargediff}
\ee
The identity for the extremal case $\mu=0$ was first shown in \cite{Cremonini:2023vwf, Cremonini:2024eog} for the EMD theory. Here we give a general proof, by including the scalar moduli $\vec \lambda$. In this more general case, we regard the thermodynamic quantities as functions of
\begin{equation}
X=X(Q_i^e,\mu,\vec\lambda)\,, \qquad X=M,\ \vec\Sigma,\ T,\ S\,.
\end{equation}

The origin of the differential identity is that the theory is invariant under the constant dilaton shift accompanied by a compensating rescaling of the gauge potentials, given by \eqref{eq:dilaton-shift-main}. Under this transformation, the electric charges transform as given in \eqref{eq:dilaton-shift-charge-transformation}. Therefore, infinitesimally, we have
\begin{equation}
\delta_D\vec\lambda=\vec c\,,\qquad \delta_D Q_i^e=\frac{1}{2}(\vec a_i\cdot\vec c)Q_i^e\,,\qquad \delta_D g_{\mu\nu}=0\,,\qquad\delta_D\mu=0\,.
\end{equation}
Since the metric is unchanged, we have
\begin{equation}
\delta_D M=0\,,\qquad \delta_D T=0\,,\qquad\delta_D S=0\,.
\end{equation}
Note that in the following derivations, $\delta_D$ denotes the constant dilaton variation whilst $\delta$ is a parameter defined in \eqref{eq:deltakappa}.

\subsubsection{Derivation from mass invariance}

In this derivation, we start with the invariance of mass under the constant dilaton shift. The invariance of the mass gives
\begin{align}
0=\delta_D M &=\sum_i
  \left(\frac{\partial M}{\partial Q_i^e}\right)_{\mu,\vec\lambda}
  \delta_D Q_i^e +
  \sum_I \left(\frac{\partial M}{\partial \lambda_I}\right)_{\mu,Q^e}
  \delta_D\lambda_I \nonumber\\
  &= \vec c\cdot \left[\frac{1}{2}\sum_i\vec a_i Q_i^e
    \left(\frac{\partial M}{\partial Q_i^e}\right)_{\mu,\vec\lambda} +
    \left(\frac{\partial M}{\partial \vec\lambda}\right)_{\mu,Q^e}\right].
\end{align}
Note that we use indices $(i,I)$ to label the quantities associated with the Maxwell fields and dilaton scalars respectively. Since $\vec c$ is arbitrary, the above implies
\begin{equation}
  \left(\frac{\partial M}{\partial \vec\lambda}\right)_{\mu,Q^e} =
  -\frac{1}{2} \sum_i \vec a_i Q_i^e \left(\frac{\partial M}{\partial Q_i^e}\right)_{\mu,\vec\lambda}\,.
  \label{eq:Mlambda_shift}
\end{equation}
Evaluating the first law \eqref{eq:extended-first-law} with basic variables $(Q_i^e,\mu,\vec\lambda)$, we have
\begin{equation}
\left(\frac{\partial M}{\partial \lambda_I}\right)_{\mu,Q^e}
=T\left(\frac{\partial S}{\partial \lambda_I}\right)_{\mu,Q^e}-\Sigma_I\,,\qquad
\left(\frac{\partial M}{\partial \mu}\right)_{Q^e,\vec\lambda}=T\left(\frac{\partial S}{\partial \mu}\right)_{Q^e,\vec\lambda}\,.
  \label{eq:firstlaw_partials}
\end{equation}
The integrability condition $\frac{\partial^2 M}{\partial\lambda_I\partial\mu}=
  \frac{\partial^2 M}{\partial\mu\partial\lambda_I}$ therefore gives
\begin{equation}
\left(\frac{\partial T}{\partial\mu}\right)_{\vec\lambda,Q^e}
\left(\frac{\partial S}{\partial\lambda_I}\right)_{\mu,Q^e}-
\left(\frac{\partial\Sigma_I}{\partial\mu}\right)_{\vec\lambda,Q^e}=
\left(\frac{\partial T}{\partial\lambda_I}\right)_{\mu,Q^e}
\left(\frac{\partial S}{\partial\mu}\right)_{\vec\lambda,Q^e}\,.\label{eq:mixed_partial}
\end{equation}
From the asymptotic-horizon relation \eqref{eq:TSmu}, we obtain
\begin{equation}
\left(\frac{\partial T}{\partial\mu}\right)_{\vec\lambda,Q^e} = \frac{T}{\mu} - \frac{T}{S}
\left(\frac{\partial S}{\partial\mu}\right)_{\vec\lambda,Q^e}\,,\qquad
\left(\frac{\partial T}{\partial\lambda_I}\right)_{\mu,Q^e}=-\frac{T}{S}
  \left(\frac{\partial S}{\partial\lambda_I}\right)_{\mu,Q^e}\,.
\end{equation}
Substituting these two expressions into \eqref{eq:mixed_partial}, we obtain
\begin{equation}
T\left(\frac{\partial S}{\partial\vec\lambda}\right)_{\mu,Q^e}=\mu
\left(\frac{\partial\vec\Sigma}{\partial\mu}\right)_{Q^e,\vec\lambda}\,.
\label{eq:TSlambda_relation}
\end{equation}
Using this in \eqref{eq:firstlaw_partials}, we find
\begin{equation}
\left(\frac{\partial M}{\partial\vec\lambda}\right)_{\mu,Q^e}
=\mu\left(\frac{\partial\vec\Sigma}{\partial\mu}\right)_{Q^e,\vec\lambda}
-\vec\Sigma\,.\label{eq:Mlambda_Sigma}
\end{equation}
Finally, combining \eqref{eq:Mlambda_shift} and \eqref{eq:Mlambda_Sigma}, we obtain
\begin{equation}
\vec\Sigma=\mu \left(\frac{\partial\vec\Sigma}{\partial\mu}\right)_{Q^e,\vec\lambda}
+\frac{1}{2} \sum_i \vec a_i Q_i^e \left(\frac{\partial M}{\partial Q_i^e}\right)_{\mu,\vec\lambda}\,.\label{eq:scalarchargediff-lambda}
\end{equation}
Setting $\vec\lambda=0$ gives the scalar-charge differential equation \eqref{eq:scalarchargediff}.

\subsubsection{Derivation from entropy invariance}

We now show that the same differential identity of the scalar charge can also be derived from a new starting point $\delta_D S=0$, which implies
\begin{equation}
\sum_i\left(\frac{\partial S}{\partial Q_i^e}\right)_{\mu,\vec\lambda}\delta_D Q_i^e
+\left(\frac{\partial S}{\partial\vec\lambda}\right)_{\mu,Q^e}\cdot\delta_D\vec\lambda
=\vec c\cdot\left[\frac{1}{2}\sum_i\vec a_i Q_i^e\left(
\frac{\partial S}{\partial Q_i^e}\right)_{\mu,\vec\lambda}
+\left(\frac{\partial S}{\partial\vec\lambda}\right)_{\mu,Q^e}\right]=0\,.
\end{equation}
Since $\vec c$ is arbitrary, we obtain the vector identity
\begin{equation}
\frac{1}{2} \sum_i \vec a_i Q_i^e \left(\frac{\partial S}{\partial Q_i^e} \right)_{\mu,\vec\lambda}
= - \left(\frac{\partial S}{\partial\vec\lambda}\right)_{\mu,Q^e}\,.
  \label{eq:S_shift_ward}
\end{equation}
Multiplying \eqref{eq:S_shift_ward} by $T$ and using \eqref{eq:TSlambda_relation}, we obtain
\begin{equation}
  \frac{T}{2} \sum_i \vec a_i Q_i^e \left( \frac{\partial S}{\partial Q_i^e}
  \right)_{\mu,\vec\lambda} = - \mu \left(\frac{\partial\vec\Sigma}{\partial\mu}
  \right)_{Q^e,\vec\lambda}\,.
  \label{eq:weighted_SQ_relation}
\end{equation}
Next, we may regard the mass either as a function of $(Q_i^e,\mu,\vec\lambda)$ or as a function of $(Q_i^e,S,\vec\lambda)$. For fixed $\mu$ and $\vec\lambda$, we have
\begin{equation}
  \left( \frac{\partial M}{\partial Q_i^e}\right)_{\mu,\vec\lambda}
  = \left(\frac{\partial M}{\partial Q_i^e}
  \right)_{S,\vec\lambda} + \left(\frac{\partial M}{\partial S}
  \right)_{Q^e,\vec\lambda} \left(\frac{\partial S}{\partial Q_i^e}
  \right)_{\mu,\vec\lambda}\,.
\end{equation}
Applying the extended first law \eqref{eq:extended-first-law}, we have
\begin{equation}
\left(\frac{\partial M}{\partial Q_i^e}\right)_{S,\vec\lambda}=
\Phi_i\,,\qquad \left(\frac{\partial M}{\partial S} \right)_{Q^e,\vec\lambda}
= T\qquad \rightarrow\qquad  \left(\frac{\partial M}{\partial Q_i^e}
\right)_{\mu,\vec\lambda} =\Phi_i + T\left(\frac{\partial S}{\partial Q_i^e}
  \right)_{\mu,\vec\lambda}\,.
\end{equation}
Multiplying the last equation above by $\frac{1}{2}\vec a_i Q_i^e$ and summing over $i$, we find
\begin{align}
  \frac{1}{2} \sum_i \vec a_i Q_i^e\left(\frac{\partial M}{\partial Q_i^e}
  \right)_{\mu,\vec\lambda}
  &= \frac{1}{2} \sum_i \vec a_i Q_i^e\Phi_i + \frac{T}{2} \sum_i \vec a_i Q_i^e
  \left(\frac{\partial S}{\partial Q_i^e} \right)_{\mu,\vec\lambda} \nonumber\\
  &= \vec\Sigma - \mu\left(\frac{\partial\vec\Sigma}{\partial\mu}
  \right)_{Q^e,\vec\lambda}\,.
\end{align}
In the last step, we used the algebraic scalar-charge relation \eqref{eq:scalar-charge-potential-identity} and also \eqref{eq:weighted_SQ_relation}. We therefore recover the differential scalar-charge relation \eqref{eq:scalarchargediff-lambda}. Note that in the first method, we made use of the asymptotic-horizon relation \eqref{eq:TSmu}, whereas in the second method, we took advantage of the scalar-charge relation \eqref{eq:scalar-charge-potential-identity}.

\section{Deriving the complete set of thermodynamic quantities}
\label{sec:twomaster}

\subsection{The setup}

We have obtained a set of identities of thermodynamic quantities, both algebraic and differential, in the general EMD-like theories. Here we show that all thermodynamic quantities can be derived from these identities, without constructing either exact or numerical black hole solutions, which is the central goal of this paper. To do so, we need to specify the basic variables of the thermodynamic system.

For simplicity, we note that the scalar moduli $\vec \lambda$ is trivial owing to the constant dilaton shifting symmetry, namely
\begin{equation}
\vec\lambda\longrightarrow\vec\lambda+\vec c\,,
\qquad Q_i^e\longrightarrow
e^{\frac12\vec a_i\cdot\vec c}Q_i^e\,,\qquad \Phi_i \rightarrow e^{-\frac12\vec a_i\cdot\vec c}\Phi_i\,,
\label{eq:dilaton-shift-charge-Phi-transformation}
\end{equation}
whereas the rest of the thermodynamic quantities are invariant under this transformation. We can thus set, without loss of generality, that
\be
\vec \lambda=0\,.
\ee
It is important to note that this is a choice of representative under the constant dilaton-shift symmetry, rather than a restriction on the local radial dynamics.  In fact, if we start from arbitrary \(\vec\lambda\), we may shift the dilaton by
\(\vec c=-\vec\lambda\) and introduce the dressed electric charges
\begin{equation}
\widehat Q_i^e
\equiv
e^{-\frac12\vec a_i\cdot\vec\lambda}Q_i^e\,,\qquad \widehat \Phi_i \equiv
e^{\frac12\vec a_i\cdot\vec\lambda}\Phi_i\,.
\label{eq:dressed-charge-Phi}
\end{equation}
The mass, entropy, temperature, scalar charges, and the non-extremality
parameter $\mu$ are invariant under this change of asymptotic-modulus frame.
Consequently, our thermodynamic identities apply at arbitrary \(\vec\lambda\) after replacing \((Q_i^e, \Phi_i)\) by \((\widehat Q_i^e,\widehat \Phi_i)\).  We suppress the hats henceforth and denote these charges again by \(Q_i^e\). The first law is now simplified to
\be
dM=TdS + \Phi_i dQ_i^e\,.\label{eq:firstlaw}
\ee

From the point of view of asymptotic structure, once we set $\vec\lambda=0$, a black hole solution is specified by mass $M$ and electric charges $Q_i^e$ whilst the scalar charges $\vec \Sigma$ are functions of $(M,Q_i^e)$. In the microcanonical ensemble associated with the first law where the mass $M$ is the thermodynamic potential, the basic variables are $(S,Q_i^e)$. We thus have
\be
X=X(S, Q_i^e)\,,\qquad \hbox{for}\qquad  X=M,\ T,\ \Phi,\ \vec\Sigma,\ \mu\,.\label{eq:Xquantity}
\ee
Alternatively, instead of using $(M,Q_i^e)$ as basic parameters, we can also consider $(\mu,Q_i^e)$ as basic parameters, where $\mu\sim T S$ is not a basic thermodynamic variable in any ensemble. In this case, we have
\be
Y=Y(\mu, Q_i^e)\,,\qquad \hbox{for}\qquad  Y=M,\ T,\ \Phi,\ \vec\Sigma\,,\ S\,.\label{eq:Yquantity}
\ee
The disadvantage of using $\mu$ as a parameter is that it gives only the parametric relations between thermodynamic quantities; however, it has an advantage that in many exact solutions, these parametric relations can be analytically given.

Before proceeding, we note another important property that all the thermodynamic quantities are invariant under constant scaling according to their dimensions, owing to the fact that the Lagrangian of our general EMD-like theories do not have a dimensionful coupling constant that would violates this homogeneity condition. As a concrete example, we note that $(M, \mu, Q_i^e)$ have the same dimensions, and hence they satisfy
\be
M=c^{-1} M(c\,\mu, c\, Q_i^e)\,,
\ee
for any nonzero constant $c$. Analogous equations can be easily applied to all other thermodynamic quantities. The origin of this homogeneity is that the general EMD-like theories considered in this paper are invariant under the trombone rescaling, namely \cite{Cremmer:1997xj}
\be
g_{\mu\nu} \rightarrow \lambda^2 g_{\mu\nu}\,,\qquad A_i \rightarrow \lambda A_i\,,\label{eq:trombone}
\ee
at the level of equations of motion. Specifically, for constant $\lambda$, the action \eqref{eq:emmd-action} transforms as $I \rightarrow \lambda^{D-2}\, I $, which does not affect the equations of motion. Applying the transformation \eqref{eq:trombone} to our black hole ansatz \eqref{eq:U-ansatz}, it takes the same form if we rescale the coordinates $(t,r)\rightarrow (t,r)/\lambda$. The solution then remains exactly the same if we rescale the integration constants $(\mu, Q_i)$ appropriately.

\subsection{Thermodynamic quantities in $(S,Q_i^e)$ variables}

In order to neaten up the dimensional coefficients in the equations, we note that the entropy of the Schwarzschild black hole is given by
\be
S_{\rm sch} = \frac14 (2\mu)^{\fft{D-2}{D-3}}\, \Omega_{D-2} = \eta\, \mu^{\fft{D-2}{D-3}}\,,\label{eq:etadef}
\ee
which defines a purely numerical constant $\eta$, depending only on dimensions $D$. We can now introduce an entropy scaling variable $R=(S/\eta)^{\delta/2}$ so that we define dimensionless charge parameters $y_i$, given by
\be
y_i = \fft{Q_i^e}{\kappa R}\,.\label{eq:yidef}
\ee
The homogeneity condition implies that the mass and scalar charge vector can be expressed as
\be
M=\kappa R\, F(y_i)\,,\qquad \vec \Sigma =\frac{\sqrt{\delta}}{2}\kappa R\, \vec G(y_i)\,,
\ee
where $F$ and $\vec G$ are the dimensionless mass and charge functions to be determined respectively. The Euler operator in the charge space is ${\cal D}\equiv \sum_i y_i \frac{\partial}{\partial y_i}$. In this parametrization, $R$ is a dimensionful variable associated with the entropy and the variables $y_i$ measure the charges relative to this scale.  We now show that the derivation of all the thermodynamic quantities is reduced to determining a single dimensionless mass function $F(y_i)$. Firstly, we note that the first law \eqref{eq:firstlaw} gives
\begin{equation}
\Phi_i=\frac{\partial F}{\partial y_i}\,;\qquad  T=\frac{\kappa\delta R}{2S}
  \left(F-{\cal D}F\right)\qquad\rightarrow \qquad \mu=R\left(F-{\cal D}F\right)\,.
\end{equation}
In the last equation, we made use of the identity \eqref{eq:TSmu}. Secondly, the algebraic scalar-charge relation \eqref{eq:scalar-charge-potential-identity}, expressed in dimensionless quantities, becomes
\begin{equation}
\vec G=\frac{1}{\sqrt{\delta}}\sum_i\vec a_i y_i\frac{\partial F}{\partial y_i}\,.
  \label{eq:GfromF}
\end{equation}
Substituting all these into \eqref{eq:quadratic-force-relation}, we obtain a closed nonlinear first-order partial differential equation for $F$:
\begin{equation}
\left(F-{\cal D}F\right)^2 = F^2 +\vec G\cdot \vec G -\frac{4}{\delta}\sum_i y_i^2\,.
  \label{eq:Fpde1}
\end{equation}
We can use the Schwarzschild black hole with all charges vanishing as the boundary condition for this differential equation and the Schwarzschild boundary condition is simply
\begin{equation}
  F(0)=1\,.
\end{equation}
We refer to \eqref{eq:Fpde1} as the master equation of thermodynamics in the microcanonical ensemble with $(S,Q_i)$ as the thermodynamic variables. Once this master equation \eqref{eq:Fpde1} is solved subject to $F(0)=1$, the mass, temperature, electric potentials, scalar charges, and non-extremal parameter $\mu$ all follow from the preceding algebraic relations and derivatives. We can thus derive all the $X$ quantities in \eqref{eq:Xquantity} of a black hole without having to construct its solution. Aside from the homogeneity property, in this derivation, we mainly used the first law, together with two algebraic relations, the quadratic-force relation \eqref{eq:quadratic-force-relation} and the scalar charge identity \eqref{eq:scalar-charge-potential-identity}.

\subsection{Thermodynamic quantities in $(\mu,Q_i^e)$ variables}

In the previous subsection, we used basic variables in microcanonical ensemble to reduce the derivation of all the thermodynamic quantities to one first-order partial differential equation on the dimensionless mass function of dimensionless charges. In practice, in many explicit examples, the thermodynamic quantities, such as $Y$ in \eqref{eq:Yquantity}, are expressed parametrically in terms of basic parameters $(\mu, Q_i^e)$. This is because $Q_i^e$ arise from the linear Maxwell equation and $\mu$ is directly related to the horizon radius parameter $r_h$. Unless in the areal-radius gauge, the $r_h$ and entropy are not simply related and it is much easier to express entropy as a function of $r_h$, rather than the other way around. In order to compare the results directly to many exact solutions existing in the literature, it is advantageous to compute the thermodynamic quantities in $(\mu,Q_i^e)$ variables, as was done in \cite{Yang:2025rud,Lu:2026lkv} for EMMD theories.

Following \cite{Yang:2025rud}, we introduce the dimensionless mass $f$ and scalar charge vector $\vec g$ in terms of dimensionless electric charge variables $x_i$
\begin{equation}
  M=\kappa\mu\, f(x_i)\,,\qquad
  \vec\Sigma=\frac{\sqrt{\delta}}{2}\,\kappa\mu\,\vec g(x_i)\,,\qquad  x_i=\frac{Q_i^e}{\kappa\mu}\,.\label{eq:MSigmaQ}
\end{equation}
The parameter $\mu$ can be factored out in $(M, \vec \Sigma)$, owing to their homogeneity property. With this convention, the quadratic force relation  \eqref{eq:quadratic-force-relation} and the scalar charge differential relation respectively become
\begin{equation}
1+\frac{4}{\delta}\sum_i x_i^2 -f^2-\vec g\cdot\vec g=0\,; \qquad  \mathcal D\vec g
=\frac{1}{\sqrt{\delta}}\sum_i\vec a_i x_i\frac{\partial f}{\partial x_i}\,,
\qquad \mathcal D\equiv \sum_i x_i\frac{\partial}{\partial x_i}\,. \label{eq:fg_constraint}
\end{equation}
These are self-contained equations for $f$ and $\vec g$ and can be solved by themselves. In order to fix the boundary conditions, it is advantageous to solve the mass and scalar charge for the EMD theory when only one electric charge is turned on. In this case, the equations can be solved analytically, using the Schwarzschild black hole as the boundary condition where all charges vanish. The result reproduces the corresponding quantities of the EMD black holes where exact solutions exist \cite{Yang:2025rud,Lu:2026lkv}. We can then use the EMD result as boundary conditions to solve the partial differential equations and obtain the corresponding black hole quantities in general EMD-like theories.

Once we have obtained $f$ and $\vec g$, we can obtain the remaining thermodynamic quantities. Defining a dimensionless entropy $s$, given by
\begin{equation}
  S=\eta\,\mu^{2/\delta}\,s(x_i),
\end{equation}
where $\eta$ is given by \eqref{eq:etadef}. It then follows from the first law \eqref{eq:firstlaw}, we have
\begin{equation}
\frac{\delta}{2}\mathcal D\log s = 1-f+\mathcal D f\,,\qquad
\Phi_i=\frac{\partial f}{\partial x_i}-\frac{\delta}{2}\frac{\partial\log s}{\partial x_i}
\,.\label{eq:entropy_pde}
\end{equation}
Finally, the temperature follows straightforwardly from \eqref{eq:TSmu}. Therefore \eqref{eq:fg_constraint} gives the master equations of thermodynamics using the $(\mu, Q_i^e)$ as the basic variables.

To summarize, in both derivations using fixed-$S$ and fixed-$\mu$ approaches, the first law \eqref{eq:firstlaw}, quadratic-force relation and asymptotic-horizon relation $TS\sim \mu$ were used. In addition, the former used the algebraic scalar-charge relation, whereas the latter made use of the differential scalar-charge relation. In either approach, the complete set of thermodynamic quantities, in addition to the scalar charges, can be derived without having to construct black hole solutions, from a chosen thermodynamic master equation.

\section{Thermodynamics in perturbation and the verification}
\label{sec:perturbation}

We have so far obtained a set of relations between the thermodynamic quantities in electrically-charged black holes in general EMD-like theories. The equations of motion can be cast into one-dimensional Toda-like equations of the form \cite{Lu:1996hh, Lu:2013uia,Ivashchuk:2002ge,Ivashchuk:2013jja,Lu:2026lkv}
\be
\ddot q_i = \exp(\sum_{j} K_{ij} q_j)\,,
\ee
where $K_{ij}$ are constants related to the inner products $A_{ij}=\vec a_i \cdot \vec a_j$  of the dilaton coupling constant vectors $\vec a_i$. For suitable choices of $\vec a_i$, these equations become the Toda equations of some classical Lie groups and hence can be solved analytically. Consequently black hole thermodynamics can be analytically derived. Many examples have been tested and the validity of the approach has been confirmed \cite{Yang:2025rud,Lu:2026lkv}. However, for general $\vec a_i$, these equations are not integrable and there exist no analytical solutions. This leads to an awkward situation that if exact black hole solutions exist, we can obtain the thermodynamic quantities explicitly, without using the new approach. On the other hand, for general cases with no exact solutions, numerical calculation can be cumbersome. Furthermore the master equations of thermodynamics \eqref{eq:Fpde1} or \eqref{eq:fg_constraint} do not accept closed-form solutions for general dilaton couplings, which is of course consistent with the fact there are no analytic black hole solutions either.

\subsection{Perturbative thermodynamics without solutions}

In this subsection, we adopt a perturbative approach to compute the thermodynamic quantities order-by-order in the power expansions of small charges $Q_i^e$ for generic $\vec a_i$. We solve the master equations \eqref{eq:Fpde1} and \eqref{eq:fg_constraint} perturbatively for both $(S,Q_i^e)$ and $(\mu, Q_i^e)$ variables respectively.

{\bf The $(S,Q_i^e)$ variables}: First, we consider solving \eqref{eq:Fpde1} for small $y_i$, all of which are assumed to be in the same order. Since the electric charges appear in the metric as function of $(Q_i^e)^2$, the dimensionless mass function $F$ must be an even functions of $y_i$. Thus if we perform Taylor expansion on $F$ in terms of $y_i$ variables, we must have
\begin{equation}
F=1+F_2+F_4+F_6+\cdots= 1+\sum_{k=1}^{\infty}F_{2k},
\end{equation}
where $F_{2k}$ is a homogeneous polynomial of $y_i$ of degree $2k$, satisfying
\begin{equation}
  F_{2k}(\lambda y_i)=\lambda^{2k}F_{2k}(y_i)\,,\qquad \hbox{or equivalently}\qquad {\cal D}F_{2k}=2kF_{2k}\,,
\end{equation}
where ${\cal D}=y_i \partial_{y_i}$. It follows that we have
\begin{equation}
\vec G = \frac{1}{\sqrt{\delta}}\sum_{k=1}^{\infty}\vec B_{2k}\,,\qquad  \vec B_{2k} \equiv \sum_i\vec a_i y_i\frac{\partial F_{2k}}{\partial y_i}.
\end{equation}
The vectors $\vec B_{2k}$ collect the coupling-weighted charge derivatives at each order.
Their scalar products, namely
\begin{equation}
\vec B_{2\ell}\cdot\vec B_{2m}=\sum_{i,j} A_{ij}\, y_i y_j \frac{\partial F_{2\ell}}{\partial y_i} \frac{\partial F_{2m}}{\partial y_j}\,,
\end{equation}
are the only terms in Eq.~\eqref{eq:Fpde1} that involve the dilaton couplings. Consequently, the theory dependence is encoded entirely by the pairwise inner products $A_{ij}=\vec a_i\cdot\vec a_j$.

At the leading order, Eq.~\eqref{eq:Fpde1} gives
\begin{equation}
F_2=\frac{1}{\delta}\sum_i y_i^2\,.
  \label{eq:F2}
\end{equation}
This quadratic solution $F_2$ fixes the leading charge response.  At every subsequent $(2k)$'th order, the contribution involving $F_{2k}$ enters linearly, while the remaining terms are built from coefficients already determined at the lower orders.  To make this triangular structure explicit, we consider the order-$2k$ part of $2F{\cal D}F-\left({\cal D}F\right)^2$; it is
\begin{equation}
4kF_{2k} + 4\sum_{\ell=1}^{k-1}(k-\ell)(1-\ell)F_{2\ell}F_{2k-2\ell}\,.
\end{equation}
Equivalently, after symmetrizing the sum under $\ell\leftrightarrow k-\ell$, the coefficient $4(k-\ell)(1-\ell)$ may be replaced by $2k-4\ell(k-\ell)$. Thus we find that
the equation \eqref{eq:Fpde1} can be solved order by order using a recursion relation
\begin{equation}
  F_{2k}= -\frac{1}{4k} \left[ \sum_{\ell=1}^{k-1} \Big(2k-4\ell(k-\ell)\Big)
    F_{2\ell}F_{2k-2\ell} + \frac{1}{\delta} \sum_{\ell=1}^{k-1}
    \vec B_{2\ell}\cdot\vec B_{2k-2\ell}\right]\,, \qquad k\geq2\,.
  \label{eq:F2krecursion}
\end{equation}
Indeed, we see that at the order $2k$, the coefficient $F_{2k}$ appears linearly, whereas every term on the right-hand side is determined by the lower-order coefficients.  The recursion therefore provides a constructive algorithm for the thermodynamic expansion to arbitrary even order in the charges. Once the mass function $F$ is given, all the remaining thermodynamic quantities can also be obtained in terms of $(S, Q_i^e)$, namely
\begin{equation}
\Phi_i=\frac{\partial F}{\partial y_i}\,,\qquad   T= \frac{\kappa\delta R}{2S}
\left[1+\sum_{k=1}^{\infty}(1-2k)F_{2k}\right]\,.
\end{equation}
Furthermore, we determine the scalar charge $\vec \Sigma$ and the non-extremal parameter $\mu$:
\begin{equation}
\vec\Sigma =\frac{1}{2}\kappa R \sum_{k=1}^{\infty}\sum_i
  \vec a_i y_i\frac{\partial F_{2k}}{\partial y_i}\,,\qquad
\mu=R\left[1+\sum_{k=1}^{\infty}(1-2k)F_{2k}\right]\,.
\end{equation}
As a concrete example, we present the mass function $F$ up to and including sixth-order of the electric charges:
\begin{eqnarray}
F&=&1 +\frac{1}{\delta}\sum_i y_i^2 -\frac{1}{2\delta^3}
  \sum_{i,j}A_{ij}y_i^2y_j^2\cr
  && -\frac{1}{6\delta^4} \left(\sum_l y_l^2\right)
  \sum_{i,j}A_{ij}y_i^2y_j^2 + \frac{2}{3\delta^5}
  \sum_{i,j,k}A_{ij}A_{jk} y_i^2y_j^2y_k^2+O(y^8)\,.\label{eq:Fperturbativeresult}
\end{eqnarray}

{\bf The $(\mu,Q_i^e)$ variables}: The perturbative solutions can also be easily obtained using $(\mu,Q_i^e)$ as the basic variables. We now expand in small dimensionless charges:
\begin{equation}
f=1+\sum_{k\geq1} f_{2k}\,,\qquad \vec g=\sum_{k\geq1}\vec g_{2k}\,,
\end{equation}
where $f_{2k}$ and $\vec g_{2k}$ are homogeneous polynomials of degree $2k$ in the variables $x_i$.  It follows from the second equation in \eqref{eq:fg_constraint} that we have
\begin{equation}
\vec g_{2k}=\frac{1}{2k\sqrt{\delta}} \sum_i \vec a_i x_i\frac{\partial f_{2k}}{\partial x_i}\,.\label{eq:g_recursion}
\end{equation}
The leading order of Eq.~\eqref{eq:fg_constraint} and the above gives
\begin{equation}
f_2=\frac{2}{\delta}\sum_i x_i^2\,,\qquad
\vec g_2=\frac{2}{\delta^{3/2}}\sum_i \vec a_i x_i^2\,.
\end{equation}
At order $2k\geq4$, Eq.~\eqref{eq:fg_constraint} gives
\begin{equation}
f_{2k} = -\frac{1}{2} \sum_{\ell=1}^{k-1} \left(f_{2\ell}f_{2k-2\ell}+
\vec g_{2\ell}\cdot\vec g_{2k-2\ell}\right).
  \label{eq:f_recursion}
\end{equation}
Equations~\eqref{eq:f_recursion} and \eqref{eq:g_recursion} provide an
algebraic recursion to arbitrary order.  At each order, the dependence on the underlying EMD-like theories enters only through the pairwise inner-product data $A_{ij}=\vec a_i\cdot\vec a_j$. Once $f$ is known, all the remaining thermodynamic quantities can be obtained straightforwardly. As a concrete example, we present here the function $f$ up to and including the fourth order:
\begin{equation}
f=1+\frac{2}{\delta}\sum_i x_i^2 -\frac{2}{\delta^2}\big(\sum_i x_i^2\big)^2-\frac{2}{\delta^3}
  \sum_{i,j} A_{ij}x_i^2x_j^2
  +O(x^6)\,.
\end{equation}
The remainder of the thermodynamic quantities follows straightforwardly perturbatively, e.g.
\be
\log s=\sum_{k\ge 1} \frac{2k-1}{\delta k} f_{2k}\,,\qquad
\Phi_i = \sum_{k \ge 1} \frac{1}{2k} \frac{\partial f_{2k}}{\partial x_i}\,.
\ee

\subsection{Thermodynamics from perturbative solutions}

We now verify the perturbative thermodynamics obtained in the previous subsection, by constructing perturbative black hole solutions. The equations of motion governing the black hole solutions are given in \eqref{eq:Uphi-divergence-form}. We shall construct perturbative solutions with small charge $Q_i^e$ as order parameters. We shall also fix the asymptotic moduli
\be
U(\infty) = 0\,,\qquad \phi(\infty) =0\,.\label{eq:asymptoticbc}
\ee
As explained earlier, a nonzero asymptotic scalar modulus may be restored by replacing the physical electric charges by the corresponding shift-invariant dressed charges; this does not alter the perturbative structure derived below.

Introducing a formal bookkeeping parameter $\epsilon$ by sending all $Q_i\to\epsilon Q_i$, we can expand both $U$ and $\vec\phi$ around the Schwarzschild black hole background
\begin{equation}
U=U_0+\epsilon^2 U_2+\epsilon^4 U_4+O(\epsilon^6),
\qquad
\vec \phi=\vec\phi_0 + \epsilon^2\vec\phi_2+\epsilon^4\vec\phi_4+O(\epsilon^6)\,.
\label{eq:expansion}
\end{equation}
In principle, we can construct such perturbative solutions to an arbitrary order, with increasing complexity. For simplicity, we shall demonstrate the results up to the $O(\epsilon^6)$ order. The equations for the zeroth-order are homogeneous, and requiring the horizon regularity, together with \eqref{eq:asymptoticbc}, give
\begin{equation}
U_0=0,
\qquad
\vec\phi_0= \vec 0\,.
\end{equation}
At the quadratic order, the equations are simply
\be
U_2' = -\frac{4\delta}{(D-3)r_h^{D-3}r^{D-2}}
\sum_i Q_i^2\,,\qquad \vec \phi_2'= -\frac{8}{(D-3)r_h^{D-3}r^{D-2}}
\sum_i\vec a_iQ_i^2\,,
\ee
where $r_h$ is given by $2\mu = r_h^{D-3}$. Imposing horizon regularity and \eqref{eq:asymptoticbc}, we have
\begin{equation}
U_2(r)=
\frac{4\delta}{(D-3)^2r_h^{D-3}r^{D-3}}
\sum_iQ_i^2\,,\qquad \vec \phi_2(r)=
\frac{8}{(D-3)^2r_h^{D-3}r^{D-3}}
\sum_i\vec a_iQ_i^2\,,
\end{equation}
The fourth-order equations are more complicated, given by
\bea
U_4' &=& \frac{16\delta}{(D-3)^3r_h^{3(D-3)}}
\left[ \frac{1}{r^{D-2}} + \frac{r_h^{D-3}}{r^{2D-5}}\right] \sum_{i,j}
\left(\delta+ A_{ij}\right)Q_i^2Q_j^2\,,\nn\\
\vec\phi_4' &=& \frac{32}{(D-3)^3r_h^{3(D-3)}}
\left[ \frac{1}{r^{D-2}} + \frac{r_h^{D-3}}{r^{2D-5}}\right]
\sum_{i,j}
\vec a_i\left(\delta+ A_{ij}\right)Q_i^2Q_j^2\,.
\eea
We find that the relevant solutions are
\bea
U_4(r)&=&-\frac{16\delta}{(D-3)^4r_h^{3(D-3)}}\left[
\frac{1}{r^{D-3}} + \frac{r_h^{D-3}}{2r^{2(D-3)}}\right]
\sum_{i,j} \left(\delta+A_{ij}\right)Q_i^2Q_j^2\,,\nn\\
\vec \phi_4(r) &=& -\frac{32}{(D-3)^4r_h^{3(D-3)}}\left[
\frac{1}{r^{D-3}}+\frac{r_h^{D-3}}{2r^{2(D-3)}}\right] \sum_{i,j}
\vec  a_i\left(\delta+A_{ij}\right)Q_i^2Q_j^2\,.
\eea
We therefore obtain the charged black hole solution up to and including quartic orders of charges.

We are now in the position to derive the thermodynamic quantities from the perturbative
analytic solutions. The leading falloffs of the function $U$ and $\vec \phi$ are given by
\eqref{eq:Uphi-asymptotic}, from which we can read off the mass $M$ \eqref{eq:mass-u-relation} and the scalar charge vector $\vec \Sigma$. The electric charges are defined by \eqref{eq:electriccharge}. We first derive the thermodynamic quantities using the $(\mu, Q_i^e)$ as variables. In this case, we introduced the dimensionless mass $f$ and scalar charge vector $\vec g$ in terms of dimensionless electric charge variables $x_i$, defined in \eqref{eq:MSigmaQ}. It follows from the leading falloff of $U$, given in \eqref{eq:Uphi-asymptotic}, we have
\bea
u_2 &=& \frac{2\delta}{(D-3)^2\mu}\sum_iQ_i^2 =\frac{2\mu}{\delta}\, X\,,\qquad X=\sum_{i} x_i^2\,,\nn\\
u_4 &=& -\frac{2\delta}{(D-3)^4\mu^3} \sum_{i,j}(\delta+A_{ij})Q_i^2Q_j^2 = -\frac{2\mu}{\delta^2}X^2 -\frac{2\mu}{\delta^3} \sum_{i,j}A_{ij}x_i^2x_j^2\,.
\eea
Using \eqref{eq:mass-u-relation} and \eqref{eq:MSigmaQ}, we have
\begin{equation}
f = 1+\frac{2}{\delta}X -\frac{2}{\delta^2}X^2 -\frac{2}{\delta^3}
\sum_{i,j}A_{ij}x_i^2x_j^2
+O(x^6)\,.\label{eq:fmuresult}
\end{equation}
It is also straightforward to read off the scalar charge vector $\vec \Sigma $ from the perturbative solution. In terms of the dimensionless charges $x_i$, we have
\be
\vec g =
\frac{2}{\delta^{3/2}}\sum_i\vec a_i x_i^2-\frac{2}{\delta^{5/2}} X\sum_i\vec a_i x_i^2
-\frac{2}{\delta^{7/2}} \sum_{i,j} \vec a_i A_{ij} x_i^2x_j^2 +O(x^6)\,.
\ee
We therefore reproduce precisely the mass and scalar charge vector functions obtained earlier without using black hole solutions. The remaining thermodynamic quantities can also be readily reproduced.

We now consider thermodynamic quantities written in $(S, Q_i^e)$ variables. It follows from the entropy formula in \eqref{eq:TSdef}, and \eqref{eq:etadef} and the definition of $R$ below, we have
\begin{equation}
R=\mu e^{U(r_h)}\,.\label{eq:RmuUh}
\end{equation}
For the perturbative solution, we have
\begin{equation}
U_2(r_h)=\frac{X}{\delta}\,,\qquad U_4(r_h) = -\frac{3}{2\delta^2}X^2
-\frac{3}{2\delta^3} \sum_{i,j}A_{ij}x_i^2x_j^2\,.
\end{equation}
Expanding \eqref{eq:RmuUh} up to the sixth-order therefore yields
\begin{equation}
\frac{R}{\mu}
=1+\frac{X}{\delta}-\frac{X^2}{\delta^2}-\frac{3}{2\delta^3} \sum_{i,j}A_{ij}x_i^2x_j^2
+O(x^6)\,.\label{eq:RmuX}
\end{equation}
It follows from $M=\kappa \mu f=\kappa R F$, we have $F=\mu f/R$, which yields
\begin{equation}
F=1+\frac{X}{\delta} -\frac{2X^2}{\delta^2} -\frac{1}{2\delta^3}
\sum_{i,j}A_{ij}x_i^2x_j^2
+O(x^6)\,.
\label{eq:Finx}
\end{equation}
Recall that in this case, $F$ is a function of dimensionless charge parameters $y_i$, defined by \eqref{eq:yidef}. The $y_i$ and $x_i$ are related by
\begin{equation}
y_i=x_i\left(1-\frac{X}{\delta}\right)+O(x^5)\,,\qquad
\sum_i y_i^2 = X-\frac{2X^2}{\delta}+O(x^6)\,.
\end{equation}
Therefore, we have
\begin{equation}
X = \sum_i y_i^2 +\frac{2}{\delta}\left(\sum_i y_i^2\right)^2 +O(y^6)\,,
\label{eq:XfromY}
\end{equation}
whereas the quartic Gram contraction is unchanged at this order:
\begin{equation}
\sum_{i,j}A_{ij}x_i^2x_j^2 =
\sum_{i,j}A_{ij}y_i^2y_j^2+O(y^6)\,.
\end{equation}
Substituting these into \eqref{eq:Finx} gives
\begin{equation}
F(y)= 1+\frac{1}{\delta}\sum_i y_i^2 -\frac{1}{2\delta^3} \sum_{i,j}A_{ij}y_i^2y_j^2
+O(y^6)\,.
\label{eq:FfixedSderived}
\end{equation}
The apparent extra term proportional to $X^2$ in \eqref{eq:Finx} cancels exactly against the quartic correction generated by the change of charge variables in \eqref{eq:XfromY}.  Equation \eqref{eq:FfixedSderived} is therefore identical to the recursion result \eqref{eq:Fperturbativeresult} up to the sixth order. The remaining thermodynamic quantities can also be obtained up to this order and we shall not present them in detail. The dimensionless scalar charge vector $\vec G$ is related to $\vec g$ by $ \vec G=\frac{\mu}{R}\vec g$. Using the same $x_i$ and $y_i$ relation, we find
\begin{equation}
\vec G =
\frac{2}{\delta^{3/2}} \sum_i\vec a_i y_i^2 -
\frac{2}{\delta^{7/2}}
\sum_{i,j}A_{ij}\vec a_i y_i^2y_j^2
+O(y^6)\,.
\end{equation}
It can be readily verified that this $\vec G$ indeed satisfies \eqref{eq:GfromF} up to this order, derived without using black hole solutions.

\section{A numerical verification}
\label{sec:numerical}

In the previous section, we have considered perturbative approach to the black hole thermodynamics in power series of small charges. In order to verify our formalism for a generic case which has no exact black hole solutions, one has to use a numerical approach, in which case, the charge configurations can be either small or large. However, there are also severe limitations in the numerical approach as it can only be performed on a case-by-case basis. As a representative example, we consider a six-dimensional theory with five Maxwell fields and two dilaton scalars. We choose the five dilaton-coupling vectors to be
\begin{equation}
\vec a_1=(1,0)\,,\quad  \vec a_2 =(0,1)\,,\quad  \vec a_3=\left(-1,\frac43\right)\,,\quad
\vec a_4=\left(\frac12,-\frac12\right)\,,\quad  \vec a_5=\left(-\frac12,-\frac12\right)\,.
\label{eq:numerical-couplings}
\end{equation}
The non-extremality parameter is fixed to $\mu=1$, and the electric charge parameters are chosen to vary along a specific one-parameter ray
\begin{equation}
\bigl(Q_1,Q_2,Q_3,Q_4,Q_5\bigr)
=\epsilon\left(1,\frac12,\frac13,\frac14,\frac15\right)\,,
\label{eq:numerical-ansatz-charge-ray}
\end{equation}
where $\epsilon$ is an overall charge-scale parameter. Again these parameters are all chosen randomly without any hidden intended implications. In six dimensions, we have $\Omega_4=8\pi^2/3$. The physical electric charges are
\begin{equation}
\bigl(Q_1^e,Q_2^e,Q_3^e,Q_4^e,Q_5^e\bigr)
=\frac{\Omega_4}{4\pi}\bigl(Q_1,Q_2,Q_3,Q_4,Q_5\bigr)
=\epsilon\,\frac{2\pi}{3}
\left(1,\frac12,\frac13,\frac14,\frac15\right).
\label{eq:numerical-charge-ray}
\end{equation}
We now construct numerical black hole solutions parametrized by $\epsilon$ and extract the corresponding thermodynamic quantities.

The relevant equations are given in \eqref{eq:Uphi-eom}, together with the Hamiltonian constraint \eqref{eq:hamiltonian-U}. As we have discussed, the regularity condition on the horizon implies that the vanishing of the Hamiltonian is automatically satisfied. Therefore, we only need to solve the second-order differential equations in \eqref{eq:Uphi-eom}. We solve the radial field equations as a
boundary-value problem (BVP) using second-order finite differences.
Specifically, the functions $U$ and $\vec\phi$ that need to be solved run from horizon $r=r_h$ to asymptotic $r=\infty$. Our blackening function $h$ naturally brings the semi-infinite range to the finite range
\be
r\in [0,\infty]\qquad \rightarrow\qquad h \in [0,1]\,.
\ee
Therefore, it is natural to use $h$ as our variable to solve the black hole differential equations in the finite-difference method. Specifically, the radial equations are discretized by second-order finite differences on \(h\in[10^{-6},1]\) and solved by Newton iteration, subject to horizon regularity and asymptotic flatness. We use 240 radial grid points for the small charge parameter \(0.02\leq\epsilon\leq0.40\) and 480 points for non-small charges with \(10\leq\epsilon\leq11\). We choose 54-digit working precision and accuracy and precision goals of 24 digits.

We then solve the master equations associated with \((\mu,Q)\) and \((S,Q)\) systems on the five-dimensional charge domains and only then evaluated on the one-parameter ray in \eqref{eq:numerical-charge-ray}. Thus, the ray was used to sample the completed multidimensional solutions rather than to replace the charge-space derivatives in the master equations. The solutions were constructed hierarchically from the exact one-charge boundaries through all two-, three-, four-, and five-charge subspaces. Charge-space derivatives were discretized with a second-order backward formula, with a first-order start on the first interior layer, and the resulting arrays were interpolated cubically. These calculations used 40-digit working precision.  For small charges, we also used the perturbative solutions, retaining all terms up to and including the \(\epsilon^8\) order.

To quantify the comparison in the small-charge interval \(I_A=[0.02,0.40]\), where the scalar charges vanish as \(\epsilon\to0\), we use the interval-normalized deviation,
\[
\Delta_A(q;X)={\rm max}\Big\{
\frac{|q_X(\epsilon)-q_{\mathrm{BVP}}(\epsilon)|}
{|q_{\mathrm{BVP}}(\epsilon)|}\Big|_{\epsilon\in I_A}\Big\}\,,
\]
to measure the maximum difference between thermodynamics derived from black hole solutions and that derived directly from master equations. Here, $q_{\rm BVP}$ denotes thermodynamic quantities derived from the numerical black hole solutions; whereas $q_X$ denotes the corresponding quantities derived from the master equations. For the \((M,\Sigma_1,\Sigma_2)\) quantities,  the direct \((\mu,Q)\) solution from the master equation \eqref{eq:fg_constraint} gives relative errors \((1.6\times10^{-7},3.6\times10^{-5},8.0\times10^{-5})\), whereas the direct \((S,Q)\) solution to master equation \eqref{eq:Fpde1} gives \((1.6\times10^{-5},1.9\times10^{-2},1.1\times10^{-2})\). The corresponding results from the order-\(\epsilon^8\) perturbative expansions give \((6.1\times10^{-6},1.1\times10^{-4},5.2\times10^{-5})\) and \((1.5\times10^{-7},2.1\times10^{-6},2.2\times10^{-6})\), respectively. In the larger-charge interval \(\epsilon\in I_B=[10,11]\), the perturbative aproach is no longer valid and we compare the $(M,\Sigma_1,\Sigma_2)$ results from numerical black hole solutions and solving directly the master equations. The maximum relative errors associated with \((\mu,Q)\) master equation are \((2.3\times10^{-5},1.1\times10^{-4}, 3.3\times10^{-4})\), and those associated with the \((S,Q)\) master equations are \((9.4\times10^{-5},4.3\times10^{-4},8.7\times10^{-4})\).

\begin{figure}[p]
\centering
\includegraphics[width=0.47\textwidth]{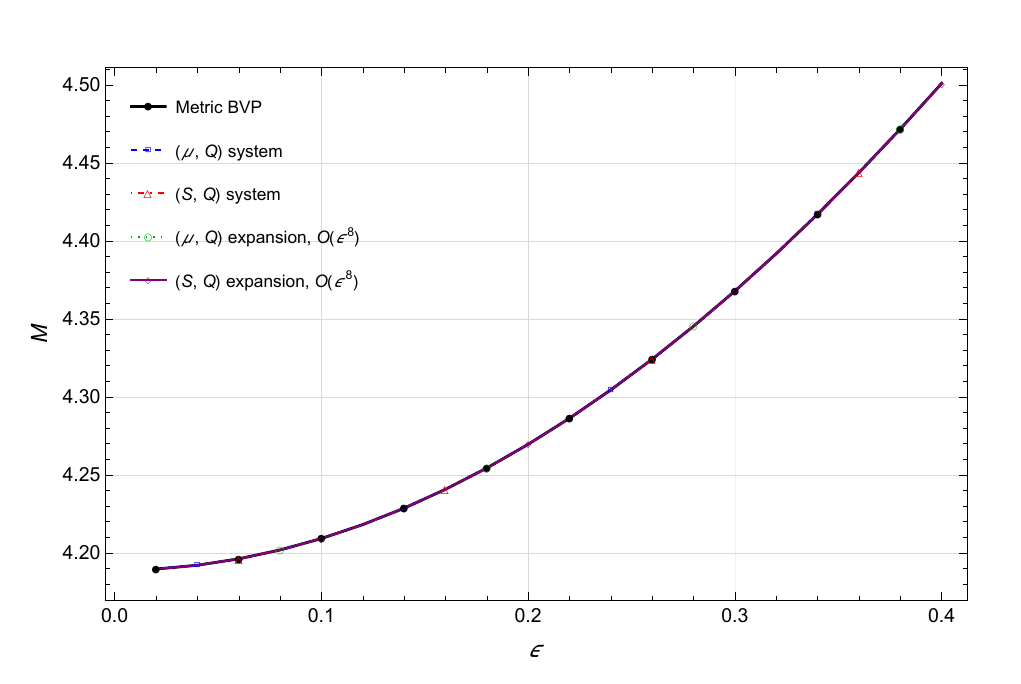}
\includegraphics[width=0.47\textwidth]{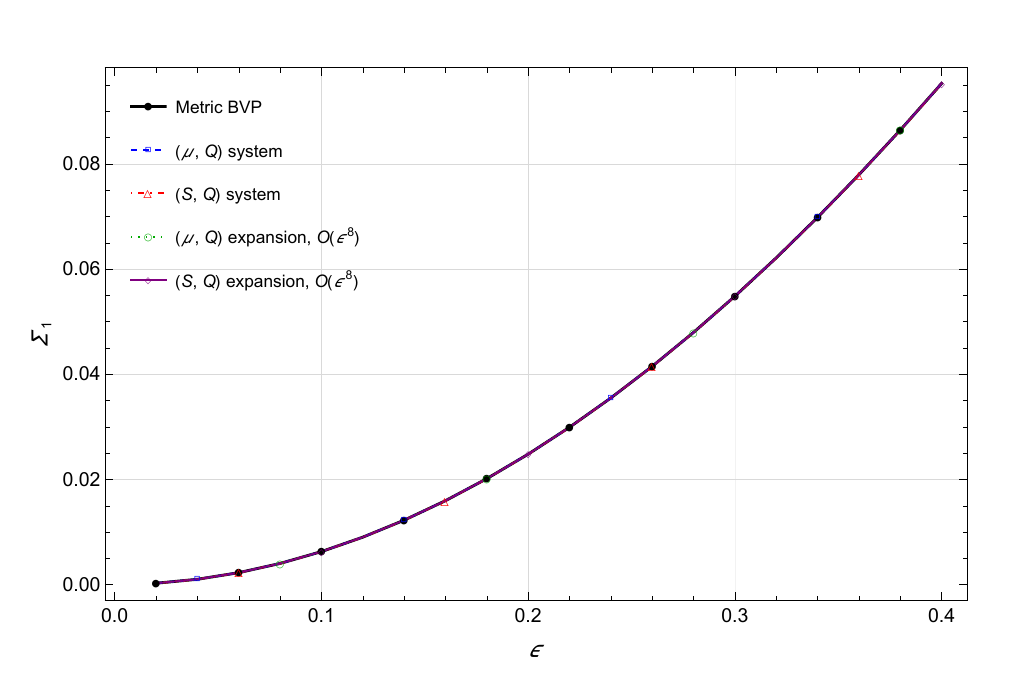}\\[2mm]
\includegraphics[width=0.47\textwidth]{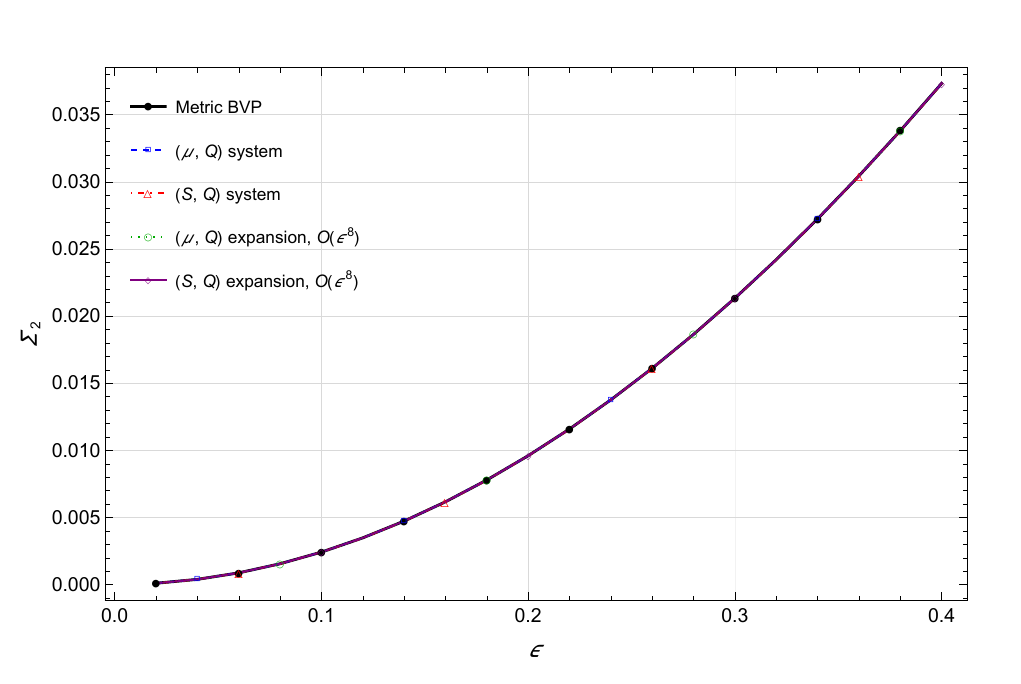}
\caption{\small Small-charge comparison for the $D=6$, $N_A=5$, $N_\phi=2$ EMD-like example specified in Eqs.~\eqref{eq:numerical-couplings}--\eqref{eq:numerical-charge-ray}.  The panels display $M$, $\Sigma_1$, and $\Sigma_2$ for $0.02\leq\epsilon\leq0.40$.  The solid black curves are extracted from the metric boundary-value solution, while the dashed and dash-dotted curves are obtained from the direct $(\mu,Q)$ and $(S,Q)$ systems, respectively.  The remaining curves show the corresponding expansions through $\mathcal O(\epsilon^8)$. These lines are all indistinguishable.}
\label{fig:numerical-small-charge}
\end{figure}

\begin{figure}[p]
\centering
\includegraphics[width=0.47\textwidth]{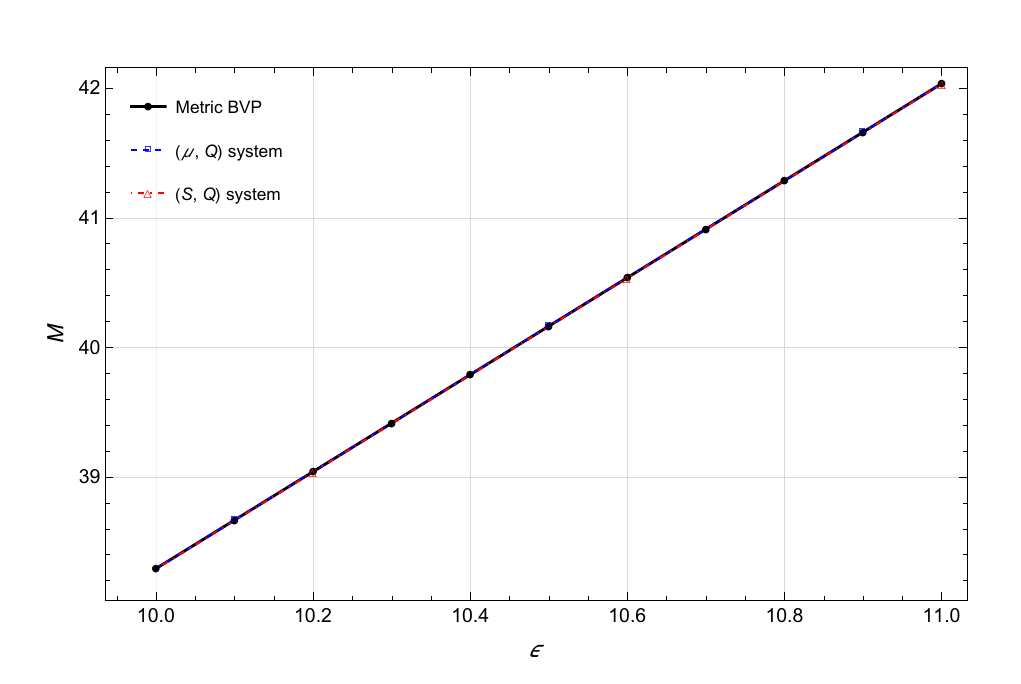}\hfill
\includegraphics[width=0.47\textwidth]{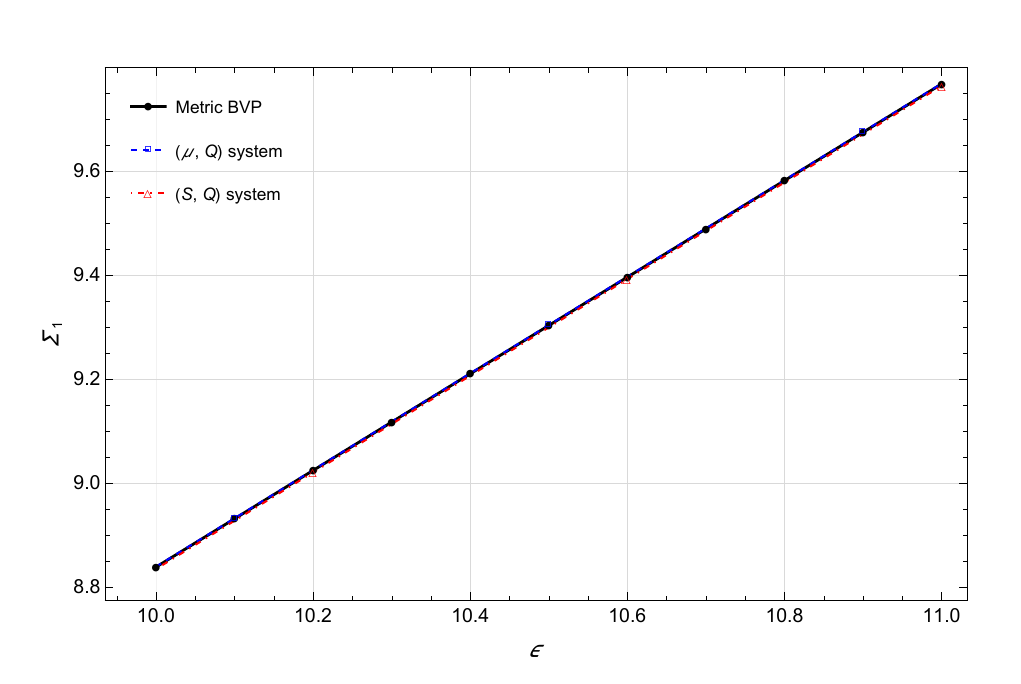}\\[2mm]
\includegraphics[width=0.47\textwidth]{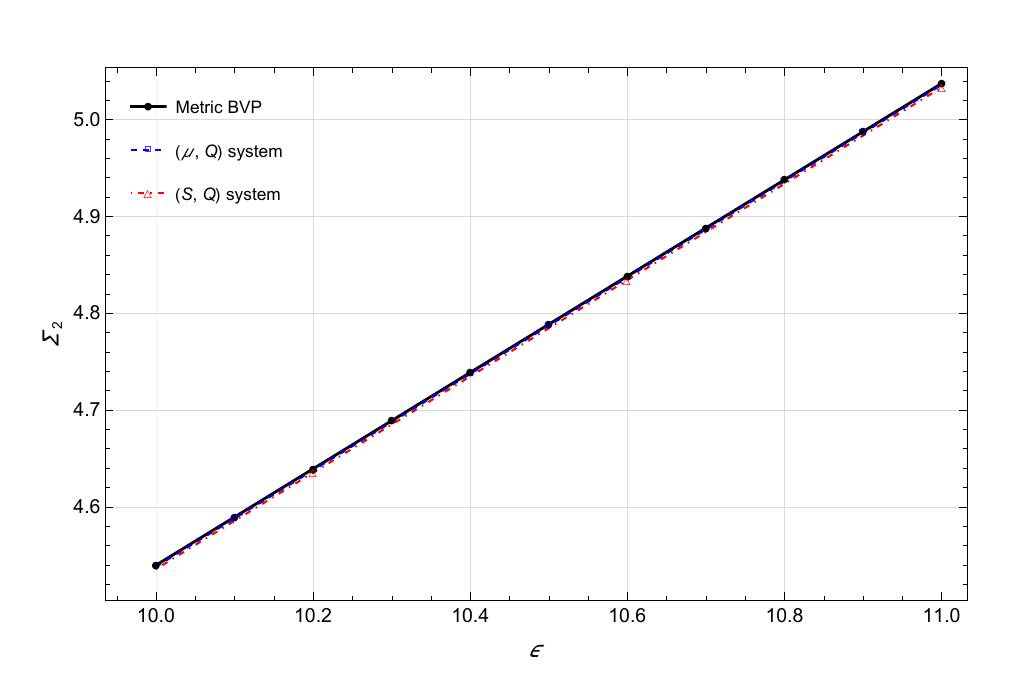}
\caption{\small Large-charge comparison for the same EMD-like example. The panels display $M$, $\Sigma_1$, and $\Sigma_2$ for $10\leq\epsilon\leq11$.  Only the metric boundary-value solution and the two nonperturbative direct thermodynamic constructions are plotted.  The figures demonstrate that the agreement persists beyond the perturbative neighborhood of the Schwarzschild point.}
\label{fig:numerical-finite-charge}
\end{figure}

The numerical results are displayed separately in the small-charge and large-charge regimes in Figures \ref{fig:numerical-small-charge} and \ref{fig:numerical-finite-charge} respectively.  The former tests the perturbative expansions around the Schwarzschild point, whereas the latter tests the two direct thermodynamic constructions at non-small charges. In both figures, ``Metric BVP'' denotes the thermodynamic data extracted from this numerical solution.  ``$(\mu,Q)$ system'' and ``$(S,Q)$ system'' denote the two direct thermodynamic constructions.  At small charge we additionally show their expansions through $\mathcal O(\epsilon^8)$.  We compare the mass and the two scalar-charge components, which test both the universal force relation and the coupling-dependent scalar sector.

Figure~\ref{fig:numerical-small-charge} shows that the entropy-fixed recursion reproduces the shared small-charge regime of the metric solution and both direct thermodynamic constructions.  Figure~\ref{fig:numerical-finite-charge} shows that the agreement of the two direct systems persists at intermediate charges. Taken together, these comparisons test the long-range-force relation and the coupling-dependent scalar-charge identities in a case without a known closed-form metric. The figures show matchings in good precisions and the lines are indistinguishable.

\section{Conclusions}
\label{sec:conclusion}

In this paper, we expanded the work of \cite{Yang:2025rud} to a general class of EMD-like theories with arbitrary numbers of both Maxwell fields and dilatonic scalars. We gave
concrete proofs to all the criteria used in the proposal of \cite{Yang:2025rud}, by deriving directly from the theories several algebraic and differential identities associated with the black hole thermodynamic quantities, scalar charges and the non-extremal parameter $\mu$. These identities allow us to determine all the thermodynamic quantities of charged static black holes without having to know the black hole solutions.

Specifically, we adopted two independent routes of derivation. One route follows the work \cite{Yang:2025rud}, where $(\mu, Q_i^e)$ are used as basic parameterizing variables for all thermodynamic quantities, as well as the scalar charges $\vec \Sigma$. The derivation of the thermodynamic quantities then reduces to the master equations in \eqref{eq:fg_constraint}. The solutions to \eqref{eq:fg_constraint} then give rise all the thermodynamic quantities. The other route is new. We used $(S,Q_i^e)$, associated with the microcanonical ensemble, as basic variables. We also obtained a corresponding master equation given in \eqref{eq:Fpde1}. The first route is advantageous for black holes that have exact solutions, in which case, thermodynamic quantities are typically expressed in terms of the electric charges $Q_i$ and non-extremal parameter $\mu$ associated directly with the horizon radius. The new route is advantageous for black holes without exact solutions, in which case, analytical expressions for thermodynamic quantities are unlikely to exist; we may as well use thermodynamic variables directly.

The master thermodynamic equations \eqref{eq:Fpde1} or \eqref{eq:fg_constraint} can be solved perturbative for the general EMD-like theory, using charges as the expansion parameters. We gave recursion relations that allowed us obtain the results to arbitrary order in expansion. We verified the results by comparing to the thermodynamic quantities obtained from the corresponding perturbative black hole solutions up to and including the quartic order in charges. We further tested the thermodynamic relations numerically in a representative six-dimensional theory along a specified charge ray.

This work essentially completed the task of deriving the thermodynamic quantities for charged spherically-symmetric and static black holes in the general string-inspired EMD theories, bypassing the constructions of black hole solutions. The robust result suggests that analogous techniques should be applicable for all the static black holes or black $p$-branes in supergravities that are low-energy effective theories of strings and M-theory, where the scalar sector is described by nonlinear sigma models, generally forming coset structures. Preliminary works can already be found in the literature \cite{Gibbons:1996af,Heidenreich:2020upe,Han:2026bim}. A bigger challenge is to generalize the technique to include rotating black holes, whose analytic solutions are much rarer. Encouraging results were obtained in \cite{huang}. All these suggest that we should try first to derive black hole thermodynamics from the theory or equations directly. Many properties, if not all, may be derived without a black hole solution.

\section*{Acknowledgement}

This work is supported in part by the National Natural Science Foundation of China (NSFC) grants No.~12375052 and No.~11935009, and also by the Tianjin University Self-Innovation Fund Extreme Basic Research Project Grant No.~2025XJ21-0007.

\end{document}